\documentclass[lettersize,journal]{IEEEtran}
\usepackage{cite}
\usepackage{url}
\usepackage{graphicx}
\usepackage{color}
\usepackage{placeins}
\usepackage{float}
\usepackage{tabularx,colortbl}
\usepackage{ifthen}
\usepackage{caption}
\usepackage{amsmath}
\usepackage{bm}
\usepackage{amssymb}
\usepackage{amsfonts}
\usepackage{xcolor}
\usepackage{titlesec}
\usepackage{tocloft}
\usepackage{tikz}
\usetikzlibrary{shapes.geometric, arrows, positioning}
\usepackage{afterpage}  
\usepackage{xcolor}
\usepackage{stfloats} 
\usepackage{comment}
\usetikzlibrary{decorations.pathreplacing}
\usepackage{tikz}
\usetikzlibrary{calc, decorations.pathreplacing}
\usepackage{graphicx}
\usepackage{subcaption}
\usepackage{caption}
\makeatletter

\newcounter{author}
\renewcommand{\author}[2][]{
   \stepcounter{author}
   \@namedef{author@\theauthor}{#2}
   \@namedef{authorlabel@\theauthor}{#1}
}

\newcounter{address}
\newcommand{\address}[2][]{
   \stepcounter{address}
   \@namedef{address@\theaddress}{#2}
   \@namedef{addresslabel@\theaddress}{#1}
}

\newcommand{\alsep}{and}

\def\newmaketitle{\par%
  \begingroup%
  \normalfont%
  \def\thefootnote{}
  \def\footnotemark{}
  \let\@makefnmark\relax
  \footnotesize
  \footnotesep 0.7\baselineskip
  \normalsize%
  \twocolumn[\thenewmaketitle\@IEEEaftertitletext]%
  \if@IEEEusingpubid
     \enlargethispage{-\@IEEEpubidpullup}%
  \fi
  \endgroup
  \setcounter{footnote}{0}\let\maketitle\relax\let\@maketitle\relax
  \gdef\@thanks{}%
  \let\thanks\relax}

\def\thenewmaketitle{
  \newpage
  \begin{center}%
    \vskip0.2em{\Huge\@IEEEcompsoconly{\sffamily}\@IEEEcompsocconfonly{\normalfont\normalsize\vskip 2\@IEEEnormalsizeunitybaselineskip
   \bfseries\large}\@title\par}\vskip1.0em\par%
    \vspace{1ex}
    \newcounter{c@author}
    \newcounter{c@tmp}
    \ifthenelse{\value{author}=2}{%
      \newcommand{\liand}{ and }}{%
      \newcommand{\liand}{, and }}
    \ifthenelse{\value{address}<2}{%
      \@nameuse{author@1}%
      \stepcounter{c@author}%
      \whiledo{\value{c@author}<\value{author}}{%
        \setcounter{c@tmp}{\value{author}}%
        \addtocounter{c@tmp}{-\value{c@author}}%
        \ifthenelse{\value{c@tmp}=1}{%
          \renewcommand{\alsep}{\liand}}{\renewcommand{\alsep}{, }}%
        \stepcounter{c@author}\alsep \@nameuse{author@\thec@author}}\\%
    }
    {
      \@nameuse{author@1}${}^{(\ref{\@nameuse{authorlabel@1}})}$%
      \stepcounter{c@author}%
      \whiledo{\value{c@author}<\value{author}}{%
      \setcounter{c@tmp}{\value{author}}%
      \addtocounter{c@tmp}{-\value{c@author}}%
      \ifthenelse{\value{c@tmp}=1}{%
        \renewcommand{\alsep}{\liand}}{\renewcommand{\alsep}{, }}%
      \stepcounter{c@author}\alsep \@nameuse{author@\thec@author}%
        ${}^{(\ref{\@nameuse{authorlabel@\thec@author}})}$%
      }
    }
    \vspace{0.2ex}

    \ifthenelse{\value{address}>0}{%
      \ifthenelse{\value{address}=1}{
        {\@nameuse{address@1}}
      }
      {
        \newcounter{c@address}

        \begin{center}
        \whiledo{\value{c@address}<\value{address}}
        {
          \refstepcounter{c@address}
            ${}^{(\thec@address)}$\,%
              \label{\@nameuse{addresslabel@\thec@address}}%
              \@nameuse{address@\thec@address}\\ %
        }
        \end{center}
      } 
    }
    {
      \relax
    }
  \end{center}
}

\makeatother

\title{Analog OFDM based on\\ Real-Time Fourier Transformation}
\author[org1]{Xiaolu Yang}
\author[org1]{Oscar C\'{e}spedes Vicente}
\author[org1]{Christophe Caloz,~\IEEEmembership{Fellow,~IEEE}}

\address[org1]{KU Leuven, Department of Electrical Engineering, Kasteelpark Arenberg 10, 3001, Leuven, Belgium}

 \newcommand{\pati}[1]{}

\begin{document}

\newmaketitle

\begin{abstract}
This paper proposes an analog orthogonal frequency division multiplexing (OFDM) architecture based on the real-time Fourier transform (RTFT). The core enabling component is a linear-chirp phaser with engineered group velocity dispersion (GVD), which realizes RTFT and performs frequency-to-time mapping in the analog domain. In this architecture, conventional digital fast Fourier transform (FFT) and inverse FFT (IFFT) processors are replaced by two linear-chirp phasers with opposite group delay dispersions, respectively. Theoretical analysis demonstrates that, under specific phaser conditions, the OFDM signal generated by the RTFT-based analog system is mathematically equivalent to that of a conventional digital OFDM system. This equivalence is further supported by simulation results, which confirm accurate symbol transmission and recovery, as well as robustness to multipath fading when a prefix is applied.
Benefiting from the use of passive microwave components, the analog OFDM system offers ultra-fast processing with reduced power consumption. Overall, this work establishes a foundation for fully analog or hybrid analog–digital OFDM systems, offering a promising solution for next-generation high-speed, wideband, and energy-efficient wireless communication platforms.
\end{abstract}

\vskip0.5\baselineskip


\section{Introduction}
\pati{Background on OFDM}
Orthogonal Frequency Division Multiplexing (OFDM) is a cornerstone in modern wireless communications~\cite{nee2000ofdm}, offering high data rates and robust transmission in multipath fading environments.
From a frequency-domain perspective, OFDM divides the entire channel into multiple narrow-band orthogonal subcarriers. Each subcarrier is sufficiently narrow such that the channel response over it can be approximated as flat fading. From a time-domain viewpoint, OFDM decomposes a high-rate data stream into multiple lower-rate parallel streams, each transmitted over different subcarriers simultaneously. The resulting extended symbol duration reduces the impact of inter-symbol interference (ISI) caused by multipath propagation. Furthermore, by appending a prefix whose length exceeds the maximum delay spread of the channel, OFDM can effectively eliminate ISI~\cite{goldsmith2005wireless}.

In conventional OFDM systems, signal multiplexing and demultiplexing are implemented using the inverse discrete Fourier transform (IDFT) at the transmitter and discrete Fourier transform (DFT) at the receiver, respectively. These are typically realized through efficient fast Fourier transform (FFT) and inverse FFT (IFFT) processors~\cite{yu2011low}. However, the growing demand for higher data rates in future wireless communications necessitates the processing of wider bandwidths, posing significant challenges for the design of FFT processors, such as larger operation latency, higher power consumption, greater hardware complexity, and elevated implementation costs~\cite{lin20051}.

To overcome these limitations, insights can be drawn from the field of ultrafast optics. In this area, the real-time Fourier transform (RTFT)~\cite{jannson1983real,muriel1999real,azana2000real,goda2013dispersive} has been extensively studied and applied for on-the-fly spectral analysis of ultra-short optical pulses. In such systems, a pulse propagates through a dispersive medium characterized by group-velocity dispersion (GVD)~\cite{saleh2008fundamentals}, which imposes frequency-dependent delays such that the temporal waveform emerging from the medium has an intensity envelope replicating the input pulse’s spectral profile. Recently, this concept has been extended to the microwave frequency domain~\cite{caloz2013analog}. The key enabling component is a linear-chirp phaser, which exhibits a group delay that varies linearly with frequency—equivalently, a quadratic phase response. When a signal passes through such a device, its spectral components emerge at different times, thereby mapping the frequency-domain content into the time domain and effectively realizing a Fourier transform in real time.

Crucially, the mapping between time and frequency can be precisely engineered by tailoring the group delay and group-delay dispersion parameters of the phaser. This capability ensures that sampling the phaser’s output in the time domain is equivalent to sampling the input signal in the frequency domain. This key property enables the implementation of the same effect as DFT operations in the analog domain, forming the basis for a novel analog realization of OFDM.

This paper presents an analog OFDM architecture based on RTFT that replaces the digital FFT/IFFT processors in conventional OFDM systems with passive phaser circuits. The proposed system enables ultra-fast operation with ultra-low power consumption, paving the way for fully analog or hybrid analog–digital OFDM implementations that can potentially overcome the limitations inherent in traditional digital FFT/IFFT-based systems.
Section~\ref{sec:Conventional OFDM} reviews the principles of conventional OFDM systems. Section~\ref{sec:OFDM-RTFT} provides the theoretical proof of the equivalence between the proposed analog OFDM and conventional OFDM under specific conditions. Section~\ref{sec:Demonstration} offers simulation-based demonstrations that validate the accuracy of signal transmission and recovery in the proposed analog OFDM system, as well as its robustness against multipath fading. Finally, Sec.~\ref{sec:Conclusion} summarizes the findings and outlines potential directions for future research.

\section{Recall of Conventional OFDM}
\label{sec:Conventional OFDM}
\pati{Key Ideas of OFDM}
OFDM divides the available bandwidth into multiple orthogonal, narrow-band sub-subcarriers, enabling efficient frequency multiplexing. Since each subcarrier occupies a narrow bandwidth, the corresponding symbol duration is extended, which inherently reduces ISI. To further combat ISI, a prefix is inserted at the beginning of each OFDM block. The prefix transforms the linear convolution with the multipath channel into a circular convolution, enabling simple frequency-domain equalization. These characteristics make OFDM a robust scheme for high-data-rate transmission in dispersive wireless environments.

\pati{Orthogonal subcarriers for Frequency Multiplexing}
The baseband OFDM signal in one OFDM block can be represented in the continuous-time\footnote{Throughout the paper, the parentheses $()$ denote time-continuous representations, while the brackets $[]$ denote for time-discrete representations.} domain as
\begin{subequations} \label{eq:OFDM_signal} 
\begin{equation} \label{eq:OFDM_signal_a_timedomain}
x(t)=\sum_{n=0}^{N-1}X[n]w(t)\varphi_n(t),
\end{equation}
where $X[n]$ denotes the $n^{\text{th}}$ symbol to be transmitted on the $n^{\text{th}}$ subcarrier, with $n=0,1,\dots, N-1$. $N$ is the number of symbols to be transmitted in one OFDM block, and thus the number of subcarriers.

The function $w(t)$ is the finite time rectangular window of duration $T_0$,
\begin{equation} \label{eq:OFDM_signal_b_timewindow}
\begin{aligned}
w(t)=\Pi\left(\frac{t-T_0/2}{T_0}\right),
\end{aligned}
\end{equation}
where $T_0 = N T_\text{s}$, with $T_\text{s}$ being the symbol duration\footnote{In the illustrative figures presented in the paper, we adopt $T_\text{s}=1 \mu\text{s}$ as an example. A detailed discussion on the values $T_\text{s}$ adopted in various OFDM standards across different frequency bands is provided later} before OFDM modulation, and $T_0$ being the duration of each data symbol on the corresponding subcarrier after OFDM modulation. Since all $N$ subcarriers are transmitted simultaneously within a single OFDM block, $T_0$ is also the duration of an OFDM block of $N$ symbols.

$\varphi_n(t)$ denotes the $n^\text{th}$ orthogonal subcarrier, given by
\begin{equation}\label{eq:OFDM_signal_c_orth_sub}
\begin{aligned}
\varphi_n(t)=A_ne^{j2\pi f_n t},
\end{aligned}
\end{equation}
where $A_n$ is the amplitude of the subcarrier. Assuming that the total transmitted power of the OFDM signal over the interval $[0, T_0]$ is equal to that of a single-carrier system with unit carrier amplitude, and that all subcarriers are equally powered, the amplitude of each subcarrier in the OFDM system is given by $A_n = 1/\sqrt{N}$. 

In Eq.~\eqref{eq:OFDM_signal_c_orth_sub}, $f_n$ denotes the subcarrier frequencies, given by
\begin{equation}\label{eq:OFDM_signal_d_subfreq}
\begin{aligned}
f_n = n/T_0,
\end{aligned}
\end{equation}
which implies that the subcarriers are uniformly spaced by $1/T_0$. This spacing ensures mutual orthogonality~\cite{couch2013digital} among the subcarriers\footnote{THe functions $\varphi_n(t)$ and $\varphi_m(t)$ are said to be orthogonal with respect to each other over the interval $a<t<b$ if they satisfy the condition $\int_a^b \varphi_n(t) \varphi^*_m(t)=0$, where $n\ne m$.} over the time interval $T_0$.
\end{subequations}

Equation~\eqref{eq:OFDM_signal_a_timedomain} can be reformulated with Eqs.~\eqref{eq:OFDM_signal_b_timewindow} and~\eqref{eq:OFDM_signal_c_orth_sub} as
\begin{equation}
x(t)=\frac{1}{\sqrt{N}}\Pi\left(\frac{t-T_0/2}{T_0}\right)\sum_{n=0}^{N-1}X[n]e^{j2\pi f_n t}.
\label{eq:OFDM_signal_timedomain}
\end{equation}

The spectrum of the OFDM signal is obtained by taking the Fourier transform\footnote{The Fourier transform~\cite{papoulis1963fourier} convention adopted in this work is defined as $X(f)=\int_{-\infty}^{\infty} x(t)e^{-j2\pi ft}dt$.} of Eq.~\eqref{eq:OFDM_signal_timedomain}, which yields
\begin{equation}\label{eq:OFDM_freq}
\begin{aligned}
\tilde{x}(f)&=\mathcal{F}\left\{x(t)\right\}\\
&=\frac{T_0}{\sqrt{N}}e^{-j2\pi f T_0/2}\sum_{n=0}^{N-1}X[n]\text{sinc}((f-f_n)T_0).
\end{aligned}
\end{equation}
This expression reveals that, in the frequency domain, the OFDM signal is composed of a weighted sum of sinc functions centered at the subcarrier frequency $f_n$. The orthogonality of the subcarriers is preserved due to the zero-crossing properties of the sinc functions at integer multiples of $1/T_0$.

\begin{figure}[h!]
    \centering
    \begin{subfigure}[t]{0.48\linewidth}
        \includegraphics[width=\linewidth]{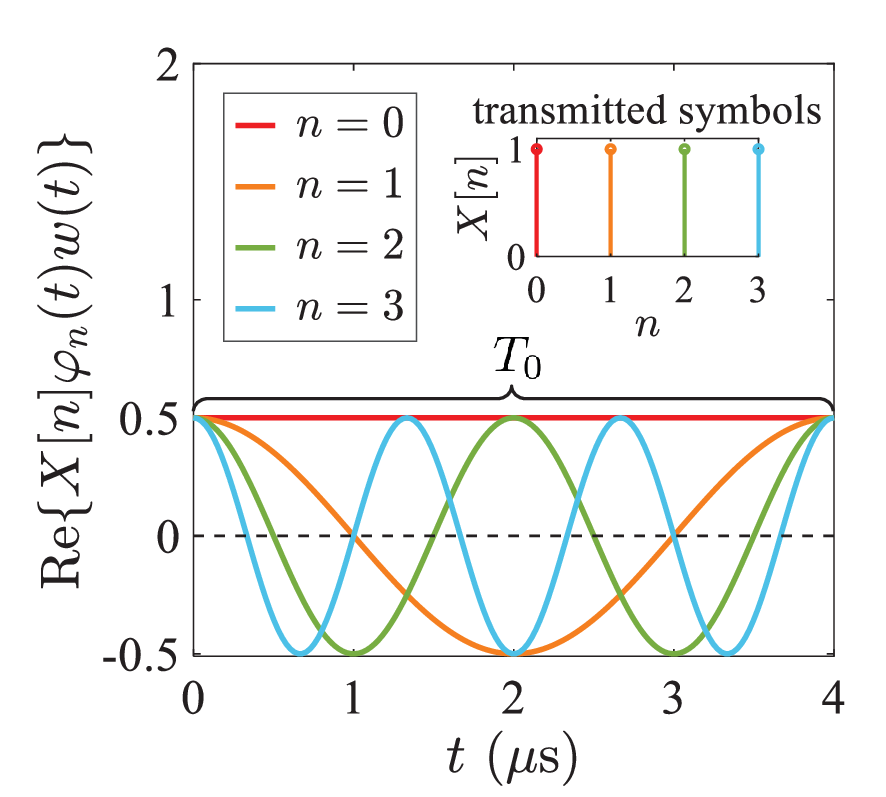}
        \captionsetup{skip=1pt}  
        \caption{}
        \label{fig:OFDM_signal_a}
    \end{subfigure}
    \hfill
    \begin{subfigure}[t]{0.48\linewidth}
        \includegraphics[width=\linewidth]{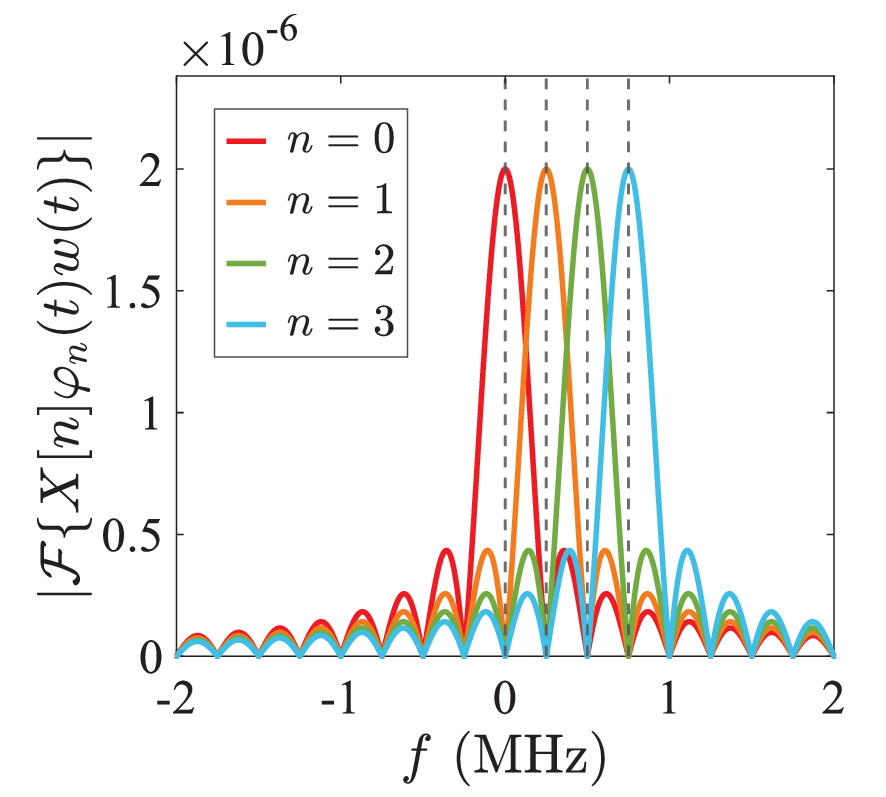}
        \captionsetup{skip=1pt}  
        \caption{}
        \label{fig:OFDM_signal_b}
    \end{subfigure}

    \vspace{1mm} 
    \begin{subfigure}[t]{0.48\linewidth}
        \includegraphics[width=\linewidth]{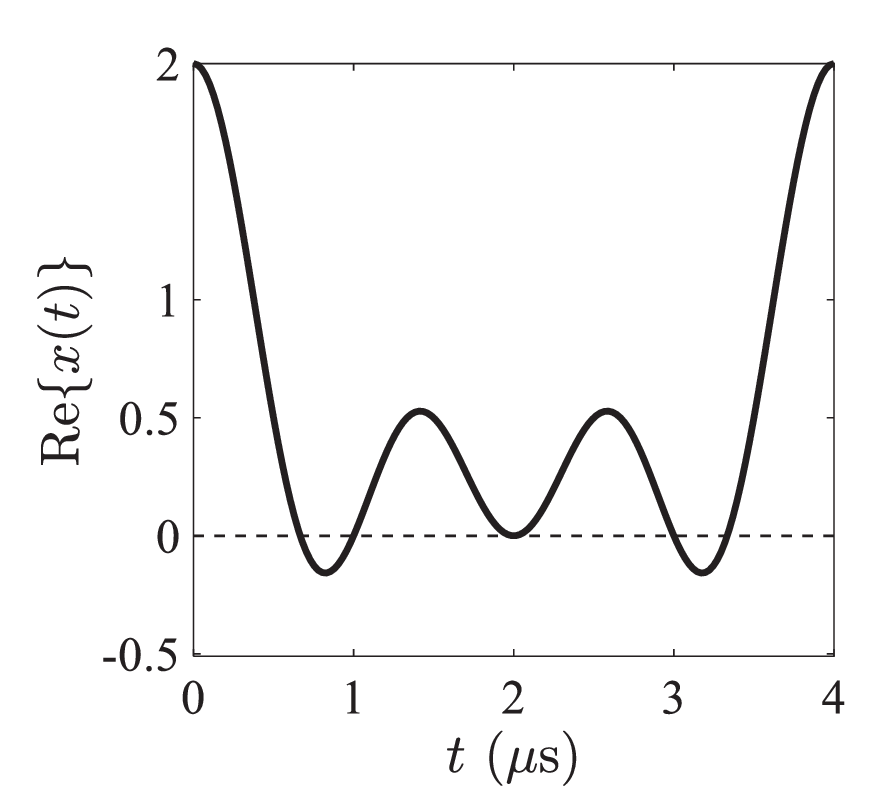}
        \captionsetup{skip=1pt}  
        \caption{}
        \label{fig:OFDM_signal_c}
    \end{subfigure}
    \hfill
    \begin{subfigure}[t]{0.48\linewidth}
        \includegraphics[width=\linewidth]{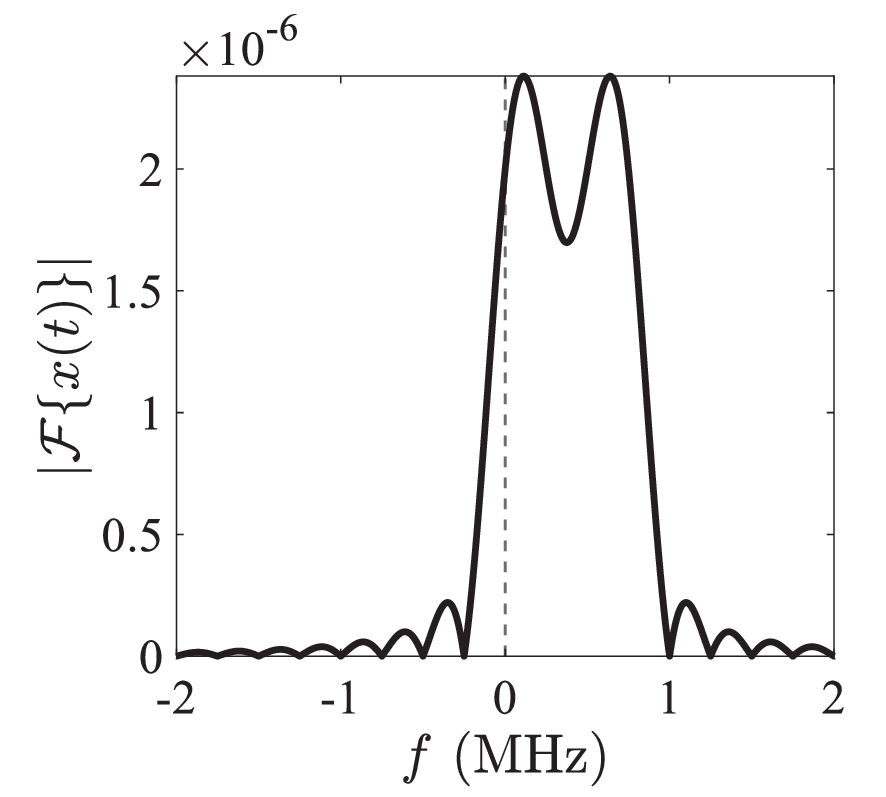}
        \captionsetup{skip=1pt}  
        \caption{}
        \label{fig:OFDM_signal_d}
    \end{subfigure}

    \caption{Illustration of an OFDM signal with $N=4$ subcarriers. 
    (a)~Real parts of the time-domain waveforms, $X[n]w(t)\varphi_n(t)$, of the four ($n=0,1,2,3$) subcarriers, $\varphi_n(t)$, modulated by the symbols, $X[n]$ (shown in the inset, with $X[n]=1$ for all $n's$ for clarity) and windowed by the rectangular function, $w(t)$ [terms of Eq.~\eqref{eq:OFDM_signal_a_timedomain} with $\varphi_n(t)$ in Eq.~\eqref{eq:OFDM_signal_c_orth_sub} and $w(t)$ in Eq.~\eqref{eq:OFDM_signal_b_timewindow}]. 
    (b)~Corresponding magnitude spectra [terms of Eq.~\eqref{eq:OFDM_freq}]. 
    (c)~Time-domain waveform (real part) of the OFDM signal [sum of the curves in (a) or Eq.~\eqref{eq:OFDM_signal_timedomain}]. 
    (d)~Corresponding magnitude spectra [sum of  curves in (b) Eq.~\eqref{eq:OFDM_freq}].}
    \label{fig:OFDM_signal}
\end{figure}

Figure~\ref{fig:OFDM_signal} illustrates an OFDM signal with $N=4$ subcarriers, where the transmitted symbols are assumed to be ${X[n]} = {1, 1, 1, 1}$, as shown in the inset of Fig.~\ref{fig:OFDM_signal}(a).
Fig.~\ref{fig:OFDM_signal}(a) depicts the time-domain waveforms $X[n] w(t) \varphi_n(t)$ for the four subcarriers ($n = 0, 1, 2, 3$), where each subcarrier $\varphi_n(t)$ is modulated by the corresponding symbol $X[n]$ and windowed by a rectangular function $w(t)$. These components correspond to the terms in Eq.~\eqref{eq:OFDM_signal_a_timedomain}, with $\varphi_n(t)$ and $w(t)$ defined in Eqs.~\eqref{eq:OFDM_signal_c_orth_sub} and \eqref{eq:OFDM_signal_b_timewindow}, respectively. Figure~\ref{fig:OFDM_signal}(b) shows the corresponding magnitude spectra of the individual subcarriers, representing the individual terms in Eq.~\eqref{eq:OFDM_freq}. Figure~\ref{fig:OFDM_signal}(c) presents the time-domain waveform of the OFDM signal, i.e., the sum of the curves in Fig.~\ref{fig:OFDM_signal}(a), consistent with  Eq.~\eqref{eq:OFDM_signal_timedomain}. Figure~\ref{fig:OFDM_signal}(d) displays the magnitude spectra of the OFDM signal, which corresponds to the sum of the individual spectra in Fig.~\ref{fig:OFDM_signal}(b), in accordance with the expression in Eq.~\eqref{eq:OFDM_freq}.

In the discrete-time form, the sampled version of the baseband OFDM signal can be obtained by substituting $t=kT_\text{s}=kT_0/N$ and Eq.~\eqref{eq:OFDM_signal_d_subfreq} into Eq.~\eqref{eq:OFDM_signal_timedomain}, which gives
\begin{equation}\label{eq:OFDM_disc}
\begin{aligned}
x\left[k\right]&=\frac{1}{\sqrt{N}}\sum_{n=0}^{N-1}X[n]e^{j2\pi \frac{n}{T_0}\frac{kT_0}{N}}\\
&=\frac{1}{\sqrt{N}}\sum_{n=0}^{N-1}X[n]e^{j2\pi \frac{kn}{N}},
\end{aligned}
\end{equation}
where $k=0,1,\dots,N-1$. In Eq.~\eqref{eq:OFDM_disc}, the window function $w(t)$ is omitted, as the finite number of symbols in the discrete-time domain inherently imposes a time window with a similar effect.

The time-discrete representation in Eq.~\eqref{eq:OFDM_disc} is essentially the IDFT of the $N$ symbols $\{X[n]\}$ in one OFDM block.  In practical systems, this transformation is efficiently implemented using an IFFT processor\footnote{The FFT/IFFT algorithm is computationally efficient when the sequence length $N$ can be factorized into smaller integers, preferably powers of 2. Under this condition, the computational complexity of the DFT is reduced from $\mathcal{O}(N^2)$ to $\mathcal{O}(N \log N)$\cite{oppenheim1999discrete}.}.

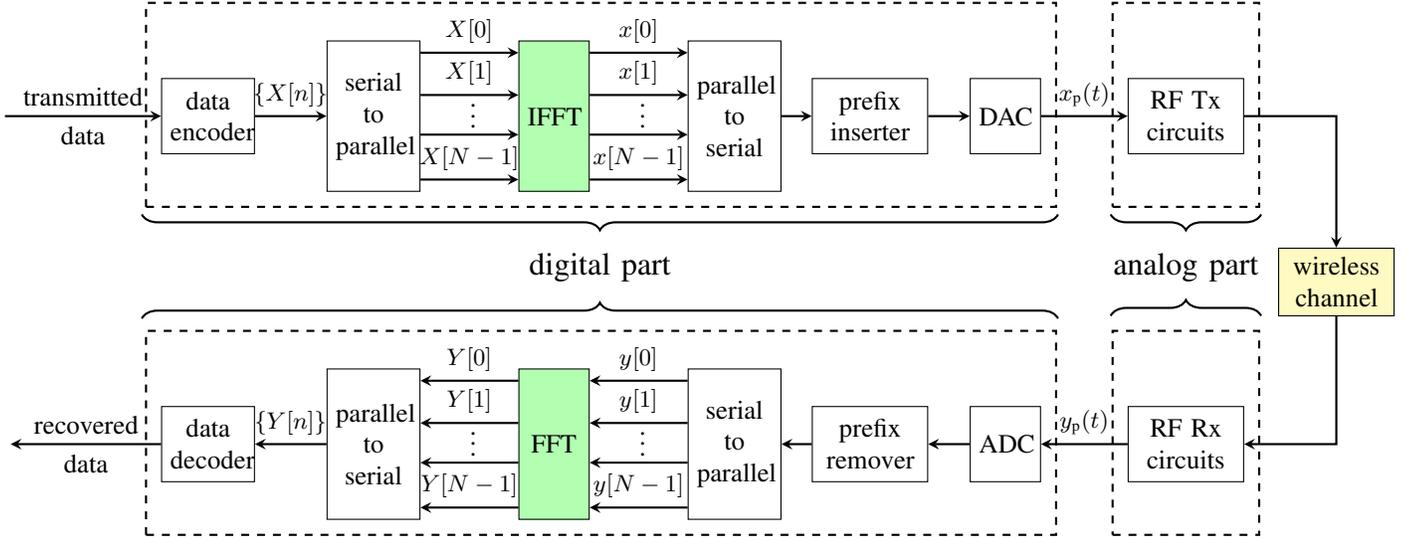
\begin{figure*}[!h]
    \centering
    \begin{tikzpicture}[node distance=1.2cm]
        \tikzstyle{block1} = [rectangle, draw, minimum width=1.3cm, text width=1.3cm, minimum height=1cm, align=center]
        \tikzstyle{block2} = [rectangle, draw, minimum width=1cm, text width=1cm, minimum height=2cm, align=center]
        \tikzstyle{block3} = [rectangle, draw, minimum width=0.7cm, text width=0.7cm, minimum height=1cm, align=center]
        \tikzstyle{block4} = [rectangle, draw, minimum width=1.2cm, text width=1cm, minimum height=1cm, align=center]
        \tikzstyle{highlight} = [rectangle, draw, minimum width=0.7cm, text width=0.7cm, minimum height=2cm, align=center, fill=green!30] 
        \tikzstyle{channel} = [rectangle, draw, minimum width=1.3cm, text width=1.3cm, align=center, fill=yellow!30]
        \tikzstyle{arrow} = [thick,->,>=stealth]
        
        \node (encoder) [block4] {data \\ encoder};
        \node (S2P_Tx) [block2, right of=encoder, xshift=1cm] {serial \\ to\\ parallel};
        \node (ifft) [highlight, right of=S2P_Tx, xshift=1.2cm] {IFFT};
        \node (P2S_Tx) [block2, right of=ifft, xshift=1.2cm] {parallel \\ to\\ serial};
        \node (addcp) [block1, right of=P2S_Tx, xshift=0.6cm] {prefix\\ inserter};
        \node (DAC) [block3, right of=addcp, xshift=0.6cm] {DAC};
        \node (rftx) [block1, right of=DAC, xshift=1.2cm] {RF Tx \\ circuits};
        \node (channel) [channel] at ([xshift=2cm, yshift=-1.7cm]rftx.south) {wireless \\ channel};
        \node (rfrx) [block1] at ([xshift=-2cm, yshift=-1.7cm]channel.south) {RF Rx \\ circuits};
        \node (ADC) [block3, left of=rfrx, xshift=-1.2cm] {ADC};
        \node (removecp) [block1, left of=ADC, xshift=-0.6cm] {prefix\\ remover};
        \node (S2P_Rx) [block2, left of=removecp, xshift=-0.6cm] {serial \\ to\\ parallel};
        \node (fft) [highlight, left of=S2P_Rx, xshift=-1.2cm] {FFT};
        \node (P2S_Rx) [block2, left of=fft, xshift=-1.2cm] {parallel \\ to\\ serial};
        \node (decoder) [block4, left of=P2S_Rx, xshift=-1cm] {data \\ decoder};
        \foreach \y/\label in {24pt/X[0], 8pt/X[1], -7pt/, -24pt/X[N-1]} {
        \draw [arrow] ([yshift=\y]S2P_Tx.east) -- ([yshift=\y]ifft.west)
        node[midway, above] {\small $\label$}; 
        }
        \node at ([xshift=20pt, yshift=3pt]S2P_Tx.east) {\(\vdots\)};
        
        \foreach \y/\label in {24pt/x[0], 8pt/x[1], -7pt/, -24pt/x[N-1]} {
        \draw [arrow] ([yshift=\y]ifft.east) -- ([yshift=\y]P2S_Tx.west)
        node[midway, above] {\small $\label$}; 
        }
        \node at ([xshift=20pt, yshift=3pt]ifft.east) {\(\vdots\)};
        
        \foreach \y/\label in {24pt/y[0], 8pt/y[1], -7pt/, -24pt/y[N-1]} {
        \draw [arrow] ([yshift=\y]S2P_Rx.west) -- ([yshift=\y]fft.east)
        node[midway, above] {\small $\label$}; 
        }
        \node at ([xshift=20pt, yshift=3pt]fft.east) {\(\vdots\)};
        
        \foreach \y/\label in {24pt/Y[0], 8pt/Y[1], -7pt/, -24pt/Y[N-1]} {
        \draw [arrow]  ([yshift=\y]fft.west) -- ([yshift=\y]P2S_Rx.east)
        node[midway, above] {\small $\label$}; 
        }
        \node at ([xshift=20pt, yshift=3pt]P2S_Rx.east) {\(\vdots\)};
        
        \draw [arrow] (-2.7cm,0) -- (encoder.west) node[midway, above] {transmitted} node[midway, below] {data};
        \draw [arrow] (encoder) -- (S2P_Tx) node[midway, above] {\small $\{X[n]\}$};
        \draw [arrow] (P2S_Tx) -- (addcp);
        \draw [arrow] (addcp) -- (DAC);
        \draw [arrow] (DAC) -- (rftx) node[midway, above] {\small $x_\text{p}(t)$};
        \draw [arrow] (rftx.east) -| (channel.north);
        \draw [arrow] (channel.south) |- (rfrx.east);
        \draw [arrow] (rfrx) -- (ADC)node[midway, above] {\small $y_\text{p}(t)$};
        \draw [arrow] (ADC) -- (removecp);
        \draw [arrow] (removecp) -- (S2P_Rx);
        \draw [arrow] (P2S_Rx) -- (decoder)node[midway, above] {\small $\{Y[n]\}$};
        \draw [arrow] (decoder.west) -- ([xshift=-2cm]decoder.west) node[midway, above] {recovered} node[midway, below] {data};
        \coordinate (TxBoxSW) at ($(encoder.south west) + (-0.2cm,-0.7cm)$);
        \coordinate (TxBoxNE) at ($(DAC.north east) + (0.2cm,1cm)$);
        \coordinate (RxBoxSW) at ($(decoder.south west) + (-0.2cm,-0.7cm)$);
        \coordinate (RxBoxNE) at ($(ADC.north east) + (0.2cm,1cm)$);
        \draw[dashed, thick] (TxBoxSW) rectangle (TxBoxNE);
        \draw[dashed, thick] (RxBoxSW) rectangle (RxBoxNE);
        \coordinate (encoder-bottom) at ($(encoder.south) + (-0.88cm,-0.8cm)$);
        \coordinate (DAC-bottom) at ($(DAC.south) + (0.7cm,-0.8cm)$);
        \coordinate (ADC-bottom) at ($(encoder.north) + (-0.88cm,-3.2cm)$);
        \coordinate (decoder-bottom) at ($(DAC.north) + (0.7cm,-3.2cm)$);
         \draw [decorate,decoration={brace, mirror, amplitude=6pt}, thick, yshift=-2pt]
        (encoder-bottom) -- (DAC-bottom) node[midway, yshift=-0.7cm, text centered] {\large \text{digital part}};
        \draw [decorate,decoration={brace, amplitude=6pt}, thick, yshift=3pt]
        (ADC-bottom) -- (decoder-bottom) ;
        
        \coordinate (TxBoxSW) at ($(rftx.south west) + (-0.2cm,-0.7cm)$);
        \coordinate (TxBoxNE) at ($(rftx.north east) + (0.2cm,1cm)$);
        \coordinate (RxBoxSW) at ($(rfrx.south west) + (-0.2cm,-0.7cm)$);
        \coordinate (RxBoxNE) at ($(rfrx.north east) + (0.2cm,1cm)$);
        \draw[dashed, thick] (TxBoxSW) rectangle (TxBoxNE);
        \draw[dashed, thick] (RxBoxSW) rectangle (RxBoxNE);
        \coordinate (rftx-leftbottom) at ($(rftx.south) + (-1cm,-0.8cm)$);
        \coordinate (rftx-rightbottom) at ($(rftx.south) + (1cm,-0.8cm)$);
        \coordinate (rfrx-lefttop) at ($(rfrx.north) + (-1cm,1.2cm)$);
        \coordinate (rfrx-righttop) at ($(rfrx.north) + (1cm,1.2cm)$);
         \draw [decorate,decoration={brace, mirror, amplitude=6pt}, thick, yshift=-2pt]
        (rftx-leftbottom) -- (rftx-rightbottom) node[midway, yshift=-0.7cm, text centered] {\large \text{analog part}};
        \draw [decorate,decoration={brace, amplitude=6pt}, thick, yshift=2pt]
        (rfrx-lefttop) -- (rfrx-righttop) ;
        
    \end{tikzpicture}
    \caption{Conventional OFDM system~\cite{goldsmith2005wireless}, based on the Fast Fourier Transform (FFT).}
    \label{fig:OFDM_system}
\end{figure*}

\pati{System Configuration based on FFT}
Figure~\ref{fig:OFDM_system} shows the block diagram of a conventional OFDM system~\cite{goldsmith2005wireless}. On the transmitter side, the input bit stream is first processed by a data encoder to generate the transmitted symbols, which are then converted from serial to parallel in each OFDM block. 
An IFFT is then applied to this parallel symbols $\{X[0], X[1], \dots, X[N-1]\}$, to  produce the corresponding OFDM symbols $\{x[0], x[1], \dots, x[N-1]\}$, which are subsequently converted back to a serial stream. To combat ISI caused by multipath propagation in wireless channels, a prefix is appended to the beginning of each OFDM block. The discrete OFDM symbols are then converted to the analog baseband OFDM signals via a digital to analog converter (DAC). This baseband signal is subsequently up-converted to the radio frequency (RF) domain using RF front-end circuits and transmitted over the wireless channel.

On the receiver side, the RF signal is down-converted back to baseband using RF reception circuits. The prefix is removed after analog to digital convertion, and the received symbols $\{y[0], y[1], \dots, y[N-1]\}$ are then  passed to an FFT block to recover the frequency-domain subcarrier components $\{Y[0], Y[1], \dots, Y[N-1]\}$. Finally, the symbol stream is fed into a data decoder to reconstruct the original transmitted information.

\setlength{\abovecaptionskip}{13pt}  

\pati{Robustness of Conventional OFDM to multipath Fading}
To gain deeper insight into how narrow-band subcarriers in OFDM help reduce ISI, we provide a detailed analysis in Appendix~\ref{sec:appendix_A_narrowband}. There, we contrast a wideband single-carrier (WSC) scheme with a single narrow-band subcarrier in OFDM. Both analytical derivations and graphical illustrations demonstrate that the extended symbol duration associated with narrow-band subcarriers significantly mitigates the effect of multipath-induced delays, thereby suppressing ISI. This comparison highlights a key advantage of OFDM in multipath fading environments.

To further combat ISI, a prefix~\cite{goldsmith2005wireless}—typically a cyclic prefix (CP), or alternatively a zero-padding prefix (ZP)—is inserted at the beginning of each OFDM block. A comprehensive discussion of prefixing techniques is provided in Appendix~\ref{sec:appendix_B_prefix}. When the prefix length exceeds the channel’s maximum delay spread, the linear convolution between the transmitted signal and the channel impulse response becomes a circular convolution over the OFDM block duration. Consequently, after removing the prefix, the received signal can be converted to the frequency domain via DFT, where each subcarrier experiences only a single complex channel coefficient. This property enables low-complexity equalization through simple division.

\section{Proposed Analog OFDM System based on RTFT}
\label{sec:OFDM-RTFT}
\subsection{OFDM System Configuration based on RTFT}
\pati {System Configuration based on RTFT}

\begin{figure*}[!h]
    \centering
    \begin{tikzpicture}[node distance=1.2cm]
        \tikzstyle{block1} = [rectangle, draw, minimum width=1.2cm, text width=1.2cm, minimum height=1cm, align=center]
        \tikzstyle{block2} = [rectangle, draw, minimum width=0.8cm, text width=0.8cm, minimum height=1cm, align=center]
        \tikzstyle{highlight} = [rectangle, draw, minimum width=1.2cm, minimum height=1cm, text width=1.2cm, align=center, fill=red!20] 
        \tikzstyle{channel} = [rectangle, draw, minimum width=1.5cm, text width=1.5cm, align=center, fill=yellow!30]
        \tikzstyle{arrow} = [thick,->,>=stealth]
        \node (encoder) [block1] {data \\ encoder};
        \node (DAC) [block2, right of=encoder, xshift=3.8cm] {DAC};
        \node (phaser_TX) [highlight, right of=DAC, xshift=1.3cm] {RT-IFT};
        \node (addcp) [block1, right of=phaser_TX, xshift=1.3cm] {prefix\\ inserter};
        \node (rftx) [block1, right of=addcp, xshift=1.3cm] {RF Tx \\ circuits};
        \node (channel) [channel] at ([xshift=2cm, yshift=-1.2cm]rftx.south) {wireless \\ channel};
        
        \node (rfrx) [block1] at ([xshift=-2cm, yshift=-1.2cm]channel.south) {RF Rx \\ circuits};
        \node (removecp) [block1, left of=rfrx, xshift=-1.3cm] {prefix\\ remover};
        \node (phaser_Rx) [highlight, left of=removecp, xshift=-1.3cm] {RT-FT};
        \node (ADC) [block2, left of=phaser_Rx, xshift=-1.3cm] {ADC};
        \node (decoder) [block1, left of=ADC, xshift=-3.8cm] {data \\ decoder};
        \draw [arrow] (-3cm,0) -- (encoder.west) node[midway, above] {transmitted} node[midway, below] {data};
        \draw [arrow] (encoder) -- (DAC) node[midway, above] {\small $\{X[n]\}$} node[midway, below] {\small $\left\{X[0], X[1], \dots, X[N-1]\right\}$};
        \draw [arrow] (DAC) -- (phaser_TX) node[midway, above] {\small $X(t)$};
        \draw [arrow] (phaser_TX) -- (addcp) node[midway, above] {\small $x(t)$};
        \draw [arrow] (addcp) -- (rftx)node[midway, above] {\small $x_\text{p}(t)$};
        
        \draw [arrow] (rftx.east) -| (channel.north);

        \draw [arrow] (channel.south) |- (rfrx.east);
        \draw [arrow] (rfrx) -- (removecp)node[midway, above] {\small $y_\text{p}(t)$};
        \draw [arrow] (removecp) -- (phaser_Rx) node[midway, above] {\small $y(t)$};
        \draw [arrow] (phaser_Rx) -- (ADC) node[midway, above] {\small $Y(t)$};
        \draw [arrow] (ADC) -- (decoder) node[midway, above] {\small $\{Y[n]\}$} node[midway, below] {\small $\left\{Y[0], Y[1], \dots, Y[N-1]\right\}$};
        \draw [arrow] (decoder.west) -- ([xshift=-2cm]decoder.west) node[midway, above] {recovered} node[midway, below] {data};
        
        \coordinate (TxBoxSW) at ($(encoder.south west) + (-0.2cm,-0.3cm)$);
        \coordinate (TxBoxNE) at ($(DAC.north east) + (0.2cm,0.4cm)$);
        \coordinate (RxBoxSW) at ($(decoder.south west) + (-0.2cm,-0.3cm)$);
        \coordinate (RxBoxNE) at ($(ADC.north east) + (0.2cm,0.4cm)$);
        \draw[dashed, thick] (TxBoxSW) rectangle (TxBoxNE);
        \draw[dashed, thick] (RxBoxSW) rectangle (RxBoxNE);
        
        \coordinate (encoder-bottom) at ($(encoder.south) + (-0.9cm,-0.5cm)$);
        \coordinate (DAC-bottom) at ($(DAC.south) + (0.7cm,-0.5cm)$);
        \coordinate (ADC-bottom) at ($(encoder.north) + (-0.9cm,-2.8cm)$);
        \coordinate (decoder-bottom) at ($(DAC.north) + (0.7cm,-2.8cm)$);
         \draw [decorate,decoration={brace, mirror, amplitude=6pt}, thick, yshift=-2pt]
        (encoder-bottom) -- (DAC-bottom) node[midway, yshift=-0.7cm, text centered] {\large \text{digital part}};
        \draw [decorate,decoration={brace, amplitude=6pt}, thick, yshift=3pt]
        (ADC-bottom) -- (decoder-bottom) ;
        
        \coordinate (TxBoxSW) at ($(phaser_TX.south west) + (-0.2cm,-0.3cm)$);
        \coordinate (TxBoxNE) at ($(rftx.north east) + (0.2cm,0.4cm)$);
        \draw[dashed, thick] (TxBoxSW) rectangle (TxBoxNE);
        \coordinate (RxBoxSW) at ($(phaser_Rx.south west) + (-0.2cm,-0.3cm)$);
        \coordinate (RxBoxNE) at ($(rfrx.north east) + (0.2cm,0.4cm)$);
        \draw[dashed, thick] (RxBoxSW) rectangle (RxBoxNE);

        \coordinate (phaser_TX-bottom) at ($(phaser_TX.south) + (-0.9cm,-0.5cm)$);
        \coordinate (rftx-bottom) at ($(rftx.south) + (0.9cm,-0.5cm)$);
        \coordinate (rfrx-bottom) at ($(phaser_TX.north) + (-0.9cm,-2.8cm)$);
        \coordinate (phaser_Rx-bottom) at ($(rftx.north) + (0.9cm,-2.8cm)$);
         \draw [decorate,decoration={brace, mirror, amplitude=6pt}, thick, yshift=-2pt]
        (phaser_TX-bottom) -- (rftx-bottom) node[midway, yshift=-0.7cm, text centered] {\large \text{analog part}};
        \draw [decorate,decoration={brace, amplitude=6pt}, thick, yshift=3pt]
        (rfrx-bottom) -- (phaser_Rx-bottom) ;
        
    \end{tikzpicture}
    \caption{Proposed OFDM system, based on the Real-Time Fourier Transform (RT-FT)~\cite{muriel1999real,caloz2013analog}.}
    \label{fig:RTFT-OFDM_system}
\end{figure*}
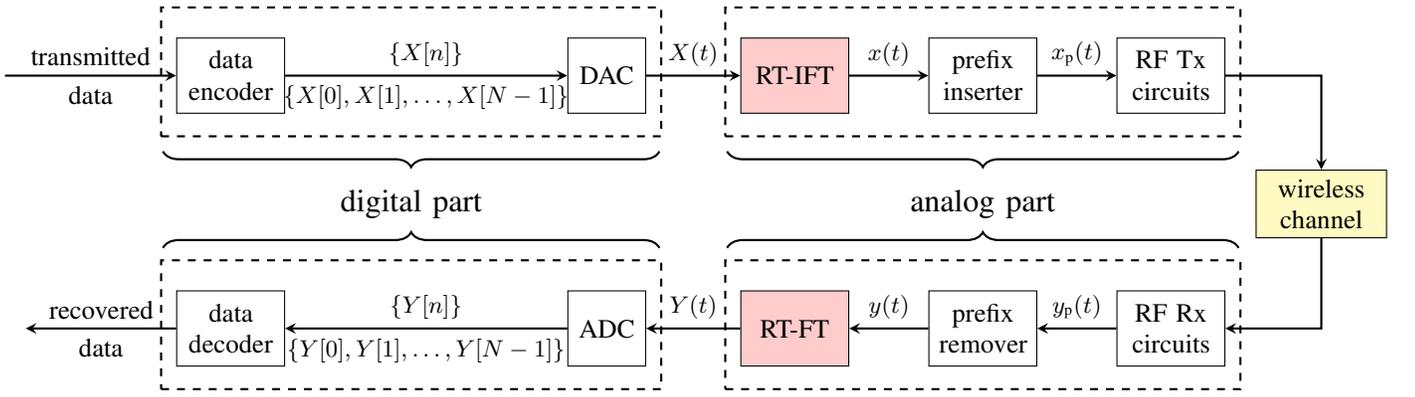

Figure~\ref{fig:RTFT-OFDM_system} illustrates the overall configuration of the proposed OFDM system, which leverages RTFT for analog signal processing. On the transmitter side, the input bit stream is first processed by a data encoder to generate the digitally modulated symbols for each OFDM block. These symbols, denoted as $\{X[n]\}$, are then converted into a continuous-time signal $X(t)$ using a DAC. This analog signal is subsequently processed by a phaser-based real-time inverse Fourier transform (RT-IFT) system, which performs an analog inverse Fourier transform via carefully engineered group delay dispersion properties. The resulting analog OFDM signal, denoted as $x(t)$, is then prepended with a prefix to mitigate ISI induced by multipath propagation. The signal with prefix, $x_\text{p}(t)$, is up-converted to the radio frequency (RF) domain and transmitted through a wireless channel.

At the receiver side, the RF signal is first down-converted to baseband using standard RF reception circuits, yielding the received signal $y_\text{p}(t)$. After removing the prefix, the signal $y(t)$ is processed by a phaser-based RTFT system, which performs a real-time Fourier transform to recover the spectral representation of the original transmitted signal. The resulting analog output $Y(t)$ is then digitized using an analog-to-digital converter (ADC), yielding a sequence of received symbols $\{Y[n]\}$. These symbols are subsequently decoded to reconstruct the original transmitted bit stream.

\setlength{\abovecaptionskip}{13pt}  

\pati {Input Signal of the RTFT-based OFDM System}

As depicted in Fig.~\ref{fig:RTFT-OFDM_system}, the input symbol sequence to the RTFT-based OFDM system is $\{X[n]\}$ for each OFDM block, which is identical to the symbol sequence in the conventional OFDM system, as shown in Fig.~\ref{fig:OFDM_system}. These digital symbols are converted into a continuous-time signal $X(t)$ through a DAC. The output of an ideal DAC can be mathematically expressed as
\begin{equation}\label{eq:DAC_output}
\begin{aligned}
X(t)&=\sum_{n=-\infty}^{+\infty}X[n]\delta(t-n T_\text{s})\\
&=\sum_{n=0}^{N-1}X[n]\delta(t-n T_\text{s}),
\end{aligned}
\end{equation}
where the second equality holds since each OFDM block contains exactly $N$ symbols. Although $X(t)$ is composed of discrete symbols, it is defined over the continuous-time domain $t \in \mathbb{R}$, and hence is treated as an analog signal\footnote{ The Dirac delta function $\delta(t - n T_\text{s})$ is a theoretical construct from continuous-time signal theory—nonzero only at isolated points—but it remains a continuous-time entity. As such, $X(t)$ represents a sequence of analog impulses, each located at $t = n T_\text{s}$, with amplitudes determined by $X[n]$, and spaced by the sampling interval $T_\text{s}$.}.

The signal $X(t)$ can be interpreted as a train of weighted Dirac impulses in continuous time, separated by the uniform sampling interval $T_\text{s}$. From the perspective of spectral analysis, such an impulse train results in a spectrum that is periodic in frequency with period $1/T_\text{s}$. The spectral content of the signal is confined to the Nyquist interval~\cite{oppenheim1999discrete}:
\begin{equation}\label{eq:frequency_range}
\begin{aligned}
f\in \left[-\frac{1}{2T_\text{s}},\frac{1}{2T_\text{s}}\right],
\end{aligned}
\end{equation}
which contains all the frequency samples of the signal before aliasing occurs.

\pati{RTFT based on GVD}
The operational principle of the RTFT block used in Fig.~\ref{fig:RTFT-OFDM_system} is detailed in Appendix~\ref{sec:appendix_C_RTFT}. Given an input signal $X(t_\text{in})$, the output of the phaser-based RTFT block~\cite{caloz2013analog} is expressed as
\begin{equation}\label{eq:RTFT_GVD}
\Psi_\text{out}(t_\text{out})=\sqrt{\frac{2\pi}{\left|\phi_2\right|}}\int_{-\infty}^{+\infty}X(t_\text{in})e^{j\frac{\phi_1+t_\text{out}}{\phi_2}t_\text{in}}dt_\text{in},
\end{equation}
where $\phi_1$ and $\phi_2$ denote the group delay and group delay dispersion parameters of the phaser, respectively. 

As observed from Eq.~\eqref{eq:RTFT_GVD}, the output signal $\Psi_\text{out}(t_\text{out})$ is essentially the Fourier transform of the input signal $X(t_\text{in})$. In particular, its time-domain profile reflects the spectral content of $X(t_\text{in})$, with the (actual) instantaneous frequency being time-dependent and given by
\begin{equation}\label{eq:RTFT_omega_t}
\omega_\text{in}(t_\text{out})=-\frac{\phi_1+t_\text{out}}{\phi_2}.
\end{equation}

In the RTFT-based OFDM system, the input signal $X(t_\text{in})$ in one OFDM block is defined as in Eq.~\eqref{eq:DAC_output}. Accordingly, the time domain of the input is restricted to the interval $t_\text{in} \in [0, T_0]$, with a sampling period of $T_\text{s}$. Based on  Eq.~\eqref{eq:frequency_range}, this temporal interval corresponds to a frequency range of $f_\text{in} \in \left[-{1}/{2T_\text{s}}, {1}/{2T_\text{s}}\right]$, which in turn translates to an angular frequency range of $\omega_\text{in} \in \left[-{\pi}/{T_\text{s}}, {\pi}/{T_\text{s}}\right]$. 

Consequently, using the time-frequency mapping relation established in  Eq.~\eqref{eq:RTFT_omega_t}, namely
$t_\text{out}(\omega_\text{in}) = -(\phi_2 \omega_\text{in} + \phi_1)$,  the resulting time domain for the output signal in one OFDM block is given by
\begin{subequations} \label{eq:t_out}
\begin{equation}\label{eq:t_out_positive_phi2}
t_\text{out}\in \left[-\frac{\pi}{T_\text{s}}\phi_2-\phi_1,~\frac{\pi}{T_\text{s}}\phi_2-\phi_1\right ],\qquad\text{if}~\phi_2\geq0,
\end{equation}
\begin{equation}\label{eq:t_out_negative_phi2}
t_\text{out}\in \left[\frac{\pi}{T_\text{s}}\phi_2-\phi_1,~-\frac{\pi}{T_\text{s}}\phi_2-\phi_1\right],\qquad\text{if}~ \phi_2\leq0.
\end{equation}
\end{subequations}

\subsection{Proof of OFDM Signal Generation via RTFT}
\pati {Proof of OFDM Signal Generation via RTFT}
 
This section provides a theoretical proof that, for a given set of transmitted symbols $\{X[n]\}$ within an OFDM block, the RTFT output is essentially equivalent to the conventional OFDM signal. 

To establish this equivalence, we reformulate the RTFT output described by Eq.~\eqref{eq:RTFT_GVD} as an expansion of complete orthogonal functions with the same structure as the conventional time-domain OFDM expression in Eq.~\eqref{eq:OFDM_signal_a_timedomain}.

For a valid block-by-block\footnote{Here, “block-by-block” refers to comparing the output signals generated from the same set of transmitted data symbols over a single OFDM block duration.} comparison, the RTFT output $\Psi_\text{out}(t_\text{out})$ must be confined to a finite time interval matching the OFDM block duration $T_0$. We therefore assume that $\Psi_\text{out}(t_\text{out})$ is nonzero only within the interval $a < t_\text{out} < a + T_0$, where $a \geq 0$.

Under this assumption, the RTFT output $\Psi_\text{out}(t_\text{out})$ over the interval $a < t_\text{out} < a + T_0$ can be expressed as~\cite{hilbert1985methods} as
\begin{equation}\label{eq:Orth_rep}
\Psi_\text{out}(t_\text{out})=\sum_{u=-\infty}^{\infty}a_u\varphi_u(t_\text{out}),
\end{equation}
where $a_u$ are the expansion coefficients and $\{\varphi_u(t_\text{out})\}$ is a complete orthogonal set.

We adopt a set of complex exponentials for the basis $\{\varphi_u(t_\text{out})\}$ in Eq.~\eqref{eq:Orth_rep}, consistently with the conventional OFDM signal in Eq.~\eqref{eq:OFDM_signal_c_orth_sub}. Specifically, we choose the basis functions
\begin{equation}\label{eq:Orth_set}
\varphi_u(t_\text{out})=A_ue^{j2\pi\frac{u}{T_0}t_\text{out}},
\end{equation}
where $A_u$ amplitudes coefficients.

The coefficients $a_u$ are obtained by inverting Eq.~\eqref{eq:Orth_rep} as
\begin{equation}\label{eq:Orth_coef_initial}
a_u=\frac{1}{K_u}\int_{a}^{a+T_0}\Psi_\text{out}(t_\text{out})\varphi^{*}_u(t_\text{out})dt_\text{out},
\end{equation}
where $K_u$ is a normalization constant following the orthogonality condition
\begin{equation}\label{eq:Ku_def}
\int_{a}^{a+T_0} \varphi_u(t_\text{out}) \varphi_v^*(t_\text{out}) \, dt_\text{out} =
\begin{cases}
    0, & u \neq v, \\
    K_u, & u = v,
\end{cases}
\end{equation}
which, upon substitution of Eq.~\eqref{eq:Orth_set} and setting $u = v$, yields
\begin{equation}\label{eq:cons_Ku}
K_u=\int_{a}^{a+T_0} \varphi_u(t_\text{out}) \varphi_u^*(t_\text{out})dt_\text{out}=A^2_uT_0.
\end{equation}

We now proceed to simplify the expression for the coefficient $a_u$ in Eq.~\eqref{eq:Orth_coef_initial}. First, based on the assumption that $\Psi_\text{out}(t_\text{out})$ is zero outside the interval $[a, a+T_0]$ of an OFDM block, Eq.~\eqref{eq:Orth_coef_initial} can be rewritten as
\begin{subequations} \label{eq:Orth_coef_further}
\begin{equation}\label{eq:Orth_coef_further_a}
\begin{aligned}
a_u=\frac{1}{K_u}\int_{-\infty}^{\infty} \Psi_\text{out}(t_\text{out})\varphi^{*}_u(t_\text{out})dt_\text{out}.
\end{aligned}
\end{equation}
Next, by successively substituting  Eqs.~\eqref{eq:cons_Ku}, \eqref{eq:RTFT_GVD} and~\eqref{eq:Orth_set} into this equation, we obtain
\begin{equation}\label{eq:Orth_coef_further_b}
\begin{aligned}
a_u&=\frac{1}{A_uT_0}\sqrt{\frac{2\pi}{\left|\phi_2\right|}}\\
&~~~~\int_{-\infty}^{\infty}\left[\int_{-\infty}^{\infty}X(t_\text{in})e^{j\frac{\phi_1+t_\text{out}}{\phi_2}t_\text{in}}dt_\text{in}\right]e^{-j2\pi\frac{u}{T_0}t_\text{out}}dt_\text{out}.
\end{aligned}
\end{equation}
This expression can be simplified as follows. First, exchanging the order of integration between $t_\text{in}$ and $t_\text{out}$ gives
\begin{equation} \label{eq:Orth_coef_further_c}
\begin{aligned}
a_u=&\frac{1}{A_u T_0}\sqrt{\frac{2\pi}{\left|\phi_2\right|}}\\
&\int_{-\infty}^{\infty}X(t_\text{in})e^{j\frac{\phi_1}{\phi_2}t_\text{in}}\left[\int_{-\infty}^{\infty}e^{j\left(\frac{t_\text{in}}{\phi_2}-2\pi\frac{u}{T_0}\right)t_\text{out}}dt_\text{out}\right]dt_\text{in}.
\end{aligned}
\end{equation}
Then, replacing the term in square brackets by its inverse Fourier transform\footnote{The inverse Fourier transform of the constant $1$ is the Dirac delta function, i.e., $\int_{-\infty}^{\infty} e^{j\omega t} dt = 2\pi \delta(\omega)$.} simplifies this last expression to
\begin{equation}\label{eq:Orth_coef_further_d}
\begin{aligned}
a_u&=\frac{1}{A_u T_0}\sqrt{\frac{2\pi}{\left|\phi_2\right|}}\\
&~~~~~\int_{-\infty}^{\infty}X(t_\text{in})e^{j\frac{\phi_1}{\phi_2}t_\text{in}}\left[2\pi\delta\left(\frac{t_\text{in}}{\phi_2}-2\pi\frac{u}{T_0}\right)\right]dt_\text{in}\\
&= \frac{2\pi}{A_u T_0}\sqrt{\frac{2\pi}{\left|\phi_2\right|}}\phi_2 X\left(\frac{2\pi u}{T_0}\phi_2\right)e^{j\frac{\phi_1}{\phi_2}\frac{2\pi u}{T_0}\phi_2},
\end{aligned}
\end{equation}
where the second equality follows from the delta function substitution rule\footnote{The delta function substitution rule is $\int_{-\infty}^{\infty}f(t)\delta(g(t))dt=\frac{f(t_0)}{g^{\prime}(t_0)}$, where $t_0$ is the root of equation $g(t)=0$~\cite{oppenheim1999discrete}.} of the Dirac delta function, ultimately giving
\begin{equation}\label{eq:Orth_coef_further_e}
\begin{aligned}
a_u&=\frac{2\pi}{A_u T_0}\sqrt{2\pi\left|\phi_2\right|}X\left(u\frac{2\pi \phi_2}{T_0}\right)e^{ju\frac{2\pi \phi_1}{T_0}}.
\end{aligned}
\end{equation}
\end{subequations}

Substituting Eqs.~\eqref{eq:Orth_coef_further_e} and~\eqref{eq:Orth_set} into Eq.~\eqref{eq:Orth_rep} provides the following expression for the RTFT output as a function of the phase parameters:
\begin{equation}\label{eq:RTFT_output_orth_rep}
\begin{aligned}
\Psi_\text{out}(t_\text{out})=&\frac{2\pi}{ T_0}\sqrt{2\pi\left|\phi_2\right|}\\
&\sum_{u=-\infty}^{\infty}X\left(u\frac{2\pi \phi_2}{T_0}\right)e^{j 2\pi u \frac{ \phi_1}{T_0}}e^{j2\pi\frac{u}{T_0}t_\text{out}},
\end{aligned}
\end{equation}
where $a\leq t_\text{out}\leq a+T_0$.

Before comparing Eq.~\eqref{eq:RTFT_output_orth_rep} with the conventional OFDM signal in Eq.~\eqref{eq:OFDM_signal_timedomain}, it is necessary to validate the assumption $a\leq t_\text{out}\leq a+T_0$ that is used in the derivation. This temporal constraint on $t_\text{out}$ can be explicitly related to the phaser parameters through Eq.~\eqref{eq:t_out}. If $\phi_2 \geq 0$, Eq.~\eqref{eq:t_out_positive_phi2} implies
\refstepcounter{equation} 
\setlength{\jot}{4pt} 
\begin{align}
-\frac{\pi}{T_\text{s}}\phi_2-\phi_1 &= a \tag{\theequation a},\label{eq:phi_positive_a} \\
\frac{\pi}{T_\text{s}}\phi_2-\phi_1 &= a+T_0 \tag{\theequation b}, \label{eq:phi_positive_b}
\end{align}
\label{eq:phi_positive}
whereas if $\phi_2 \leq 0$, Eq.~\eqref{eq:t_out_negative_phi2} implies
\refstepcounter{equation} 
\setlength{\jot}{4pt} 
\begin{align}
\frac{\pi}{T_\text{s}}\phi_2-\phi_1 &= a \tag{\theequation a},\label{eq:phi_negative_a} \\
-\frac{\pi}{T_\text{s}}\phi_2-\phi_1 &= a+T_0 \tag{\theequation b} \label{eq:phi_negative_b}
\end{align}
\label{eq:phi_negative}
Solving these two systems for $\phi_1$ and $\phi_2$ yields
\refstepcounter{equation} 
\setlength{\jot}{4pt} 
\begin{align}
\phi_2 &= \pm \frac{T_0T_\text{s}}{2\pi}=\pm \frac{NT^2_\text{s}}{2\pi} \tag{\theequation a},\label{eq:phaser_cond1_phi2} \\
\phi_1 &=-\frac{T_0}{2}-a \tag{\theequation b}, \label{eq:phaser_cond1_phi1}
\end{align}
\label{eq:phaser_cond1}
which are the phaser conditions ensuring that the assumption on $t_\text{out}$---used in the derivation of  Eq.~\eqref{eq:RTFT_output_orth_rep}---is satisfied.

We can now proceed to compare the RTFT output with the conventional OFDM signal. To facilitate this comparison, we first rewrite the conventional OFDM signal in Eq.~\eqref{eq:OFDM_signal_timedomain} with the substitution\footnote{For the sake of illustration, we assume that the sampling frequency is equal to the symbol rate, i.e., $X[n]=X(nT_\text{s})$.} $X[n]=X(nT_\text{s})$
\begin{equation}
x(t)=\frac{1}{\sqrt{N}}\sum_{n=0}^{N-1}X(nT_\text{s})e^{j2\pi \frac{n}{T_0} t},~~~~0\leq t\leq T_0.
\label{eq:OFDM_signal_continusX}
\end{equation}
We see then that an extra condition for equivalence of this expression with Eq.~\eqref{eq:RTFT_output_orth_rep} is that the two relations share the same sampling time, viz.,
\begin{subequations}
\begin{equation} \label{eq:u_condi_a}
u\frac{2\pi \phi_2}{T_0} = nT_\text{s},
\end{equation}
where the substitution of Eq.~\eqref{eq:phaser_cond1_phi2} translates into the relation
\begin{equation} \label{eq:u_condi_b}
u = \pm n,
\end{equation}
\end{subequations}
where $n=0,1,\dots,N-1$.

Substituting the constraint in Eq.~\eqref{eq:phaser_cond1_phi2} on $\phi_2$ and the constraint in Eq.~\eqref{eq:u_condi_b} on $u$ into Eq.~\eqref{eq:RTFT_output_orth_rep} reformulates the series representation of the RTFT output as\footnote{Substituting $u=n$ with $\phi_2=\frac{NT^2_\text{s}}{2\pi}$ or $u=-n$ with $\phi_2=- \frac{NT^2_\text{s}}{2\pi}$ into Eq.~\eqref{eq:RTFT_output_orth_rep} yields the same result, given by  Eq.~\eqref{eq:RTFT_output_orth_rep_1}.}
\begin{equation}\label{eq:RTFT_output_orth_rep_1}
\begin{aligned}
\Psi_\text{out}(t_\text{out})=\frac{2\pi}{ \sqrt{N}}\sum_{n=0}^{N-1}X(nT_\text{s})e^{j2\pi n \frac{\phi_1}{T_0}}  e^{j2\pi\frac{n}{T_0}t_\text{out}}.
\end{aligned}
\end{equation}

Next, we discuss the selection of the parameter $\phi_1$. According to Eq.~\eqref{eq:phaser_cond1_phi1}, $\phi_1$ is related to the starting time of the RTFT output $a$ as $a = -T_0/2 - \phi_1$. To ensure causality, this starting time must satisfy $a \geq 0$, leading to the following constraint on $\phi_1$
\begin{equation}\label{eq:phi1_condi}
\begin{aligned}
\phi_1\leq  -T_0/2.
\end{aligned}
\end{equation}
Two cases may be distinguished.

\vspace{1mm}
\noindent\textbf{Case 1:} \textit{Zero starting time}

If $a = 0$ (i.e., $t_\text{out} \in [0, T_0]$), Eq.~\eqref{eq:phaser_cond1_phi1} becomes
\begin{equation}\label{eq:phi1_value1}
\begin{aligned}
\phi_1= -T_0/2.
\end{aligned}
\end{equation}
Substituting this relation into Eq.~\eqref{eq:RTFT_output_orth_rep_1} transforms the RTFT output to
\begin{equation}\label{eq:RTFT_OFDM_orth_rep_2}
\begin{aligned}
x_\text{RTFT}(t_\text{out})=\frac{2\pi}{ \sqrt{N}}\sum_{n=0}^{N-1}X(nT_\text{s})e^{jn\pi}e^{j2\pi\frac{n}{T_0}t_\text{out}},
\end{aligned}
\end{equation}
which is essentially---ignoring the $2\pi$ factor---identical to the conventional OFDM signal in Eq.~\eqref{eq:OFDM_signal_continusX}, except for the additional phase term $e^{j n \pi}$, which, as will be shown later, can be compensated at the receiver side.

\vspace{1mm}
\noindent\textbf{Case 2:} \textit{Phase-aligned RTFT output}

\begin{equation}\label{eq:phi1_value2}
\phi_1 =-T_0.
\end{equation}
Substituting this into Eq.~\eqref{eq:phaser_cond1_phi1} gives $a = T_0/2$, implying that the RTFT output signal spans the interval $t_\text{out} \in [T_0/2, 3T_0/2]$. Using this value of $\phi_1$ in Eq.~\eqref{eq:RTFT_output_orth_rep_1}, the RTFT output becomes 
\begin{equation}\label{eq:RTFT_OFDM_orth_rep_3}
\begin{aligned}
x_\text{RTFT}(t_\text{out})=\frac{2\pi}{ \sqrt{N}}\sum_{n=0}^{N-1}X(nT_\text{s})e^{j2\pi\frac{n}{T_0}t_\text{out}},
\end{aligned}
\end{equation}
where $t_\text{out} \in [T_0/2, 3T_0/2]$, which is again identical Eq.~\eqref{eq:OFDM_signal_continusX}, this time without any phase difference term. 

\vspace{1mm}
Now, we can explicitely rewrite the integral form of the RTFT output in Eq.~\eqref{eq:RTFT_GVD} as
\begin{equation}\label{eq:OFDM_RTFT}
x_\text{RTFT}(t_\text{out})=\sqrt{\frac{2\pi}{\left|\phi_2\right|}}\int_{-\infty}^{+\infty}X(t_\text{in})e^{j\frac{\phi_1+t_\text{out}}{\phi_2}t_\text{in}}dt_\text{in},
\end{equation}
with the parameters $\phi_2$ and $\phi_1$ being given by Eqs.~\eqref{eq:phaser_cond1_phi2} and~\eqref{eq:phi1_value1} or~\eqref{eq:phi1_value2}, respectively.

\begin{figure}[h!]
    \centering
    \begin{subfigure}[t]{0.48\linewidth}
        \includegraphics[width=\linewidth]{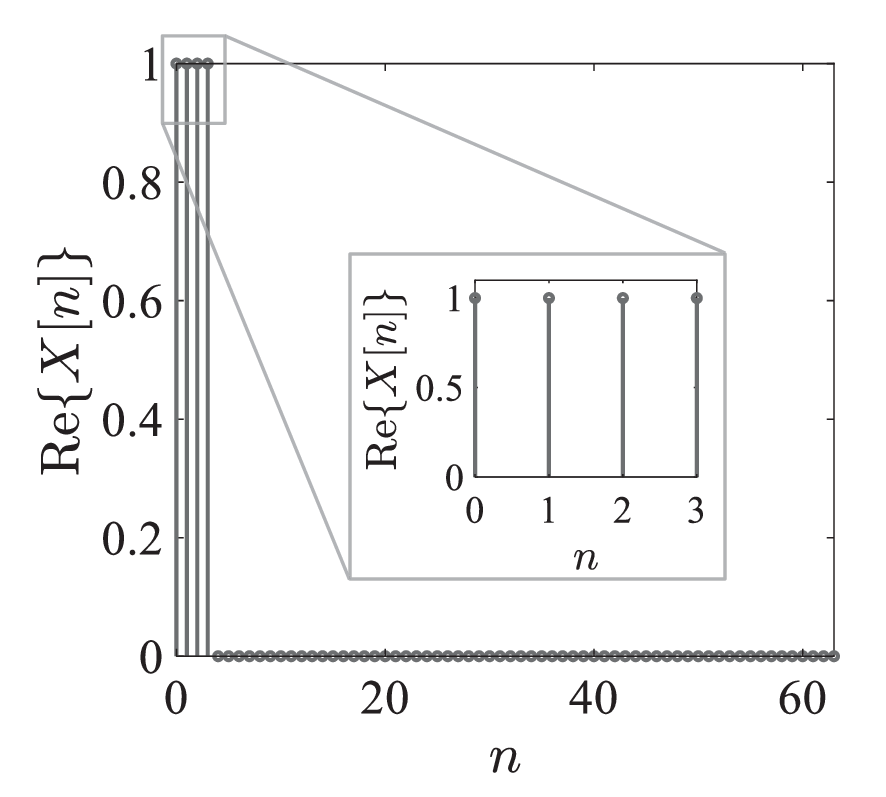}
        \captionsetup{skip=1pt}  
        \caption{}
        \label{fig:analog_symbols_a}
    \end{subfigure}
    \hfill
    \begin{subfigure}[t]{0.48\linewidth}
        \includegraphics[width=\linewidth]{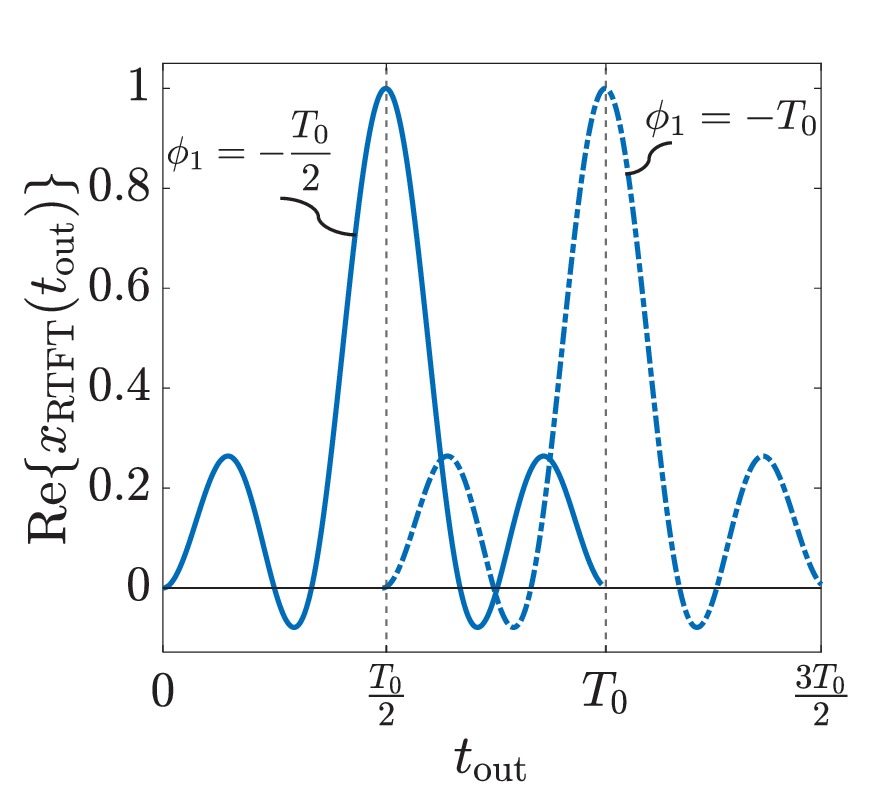}
        \captionsetup{skip=1pt}  
        \caption{}
        \label{fig:OFDM_RTFT_b}
    \end{subfigure}
    \caption{OFDM signal block generated by RTFT with $N=64$. 
    (a)~Transmitted symbols (real part), $\text{Re}\{X[n]\}$. For illustration, the transmitted symbols in the block are set to $\{X[n]\}=\{1,1,1,1,0,0,\dots,0\}$. 
    (b)~Normalized time-domain waveform (real-part) of the corresponding OFDM signal [Eq.~\eqref{eq:OFDM_RTFT}, with $\phi_2$ given by Eq.~\eqref{eq:phaser_cond1_phi2} and $\phi_1$ given by Eq.~\eqref{eq:phi1_value1} or Eq.~\eqref{eq:phi1_value2}].}
    \label{fig:OFDM_RTFT}
\end{figure}

Figure~\ref{fig:OFDM_RTFT} illustrates the OFDM signal generated via the RTFT approach. For demonstration purposes, we consider $N = 64$ and the transmitted symbol sequence within one OFDM block is chosen as $\{X[n]\} = \{1, 1, 1, 0, 0, \dots, 0\}$, as shown in Fig.~\ref{fig:OFDM_RTFT}(a). Figure~\ref{fig:OFDM_RTFT}(b) displays the normalized OFDM signal generated using the RTFT approach, obtained by evaluating Eq.~\eqref{eq:OFDM_RTFT} with $\phi_2$ set according to Eq.~\eqref{eq:phaser_cond1_phi2}. The two curves represent the two different choices of the parameter $\phi_1$: the solid curve corresponds to $\phi_1 = -T_0/2$ [Eq.~\eqref{eq:phi1_value1}], while the dashed curve corresponds to $\phi_1 = -T_0$ [Eq.~\eqref{eq:phi1_value2}]. In both cases, the RTFT output signal $x_\text{RTFT}(t_\text{out})$ spans, as it should, a time duration equal to one OFDM block, i.e., $T_0$, as enforced by the parameter $\phi_2$, and remains unaffected by the choice of $\phi_1$.

It turns out that the two distinct continuous-time curves in Fig.~\ref{fig:OFDM_RTFT}(b) will result in the same discrete-time result. Indeed, inserting $t_\text{out}=kT_\text{s}=kT_0/N$ into Eq.~\eqref{eq:RTFT_OFDM_orth_rep_2} and $t_\text{out}=T_0/2+kT_\text{s}=T_0/2+kT_0/N$ into Eq.~\eqref{eq:RTFT_OFDM_orth_rep_3} leads to the unique result
\begin{equation}\label{eq:RTFT_OFDM_disc}
\begin{aligned}
x_\text{RTFT}[k]
&= \frac{2\pi}{\sqrt{N}} \sum_{n=0}^{N-1} X[n] e^{j2\pi \frac{kn}{N}} e^{jn\pi},
\end{aligned}
\end{equation}
where $k = 0,1,\dots, N-1$.

\pati{Comparison with the Conventional OFDM Signal}

\begin{figure}[h!]
    \centering
    \begin{subfigure}[t]{0.48\linewidth}
        \includegraphics[width=\linewidth]{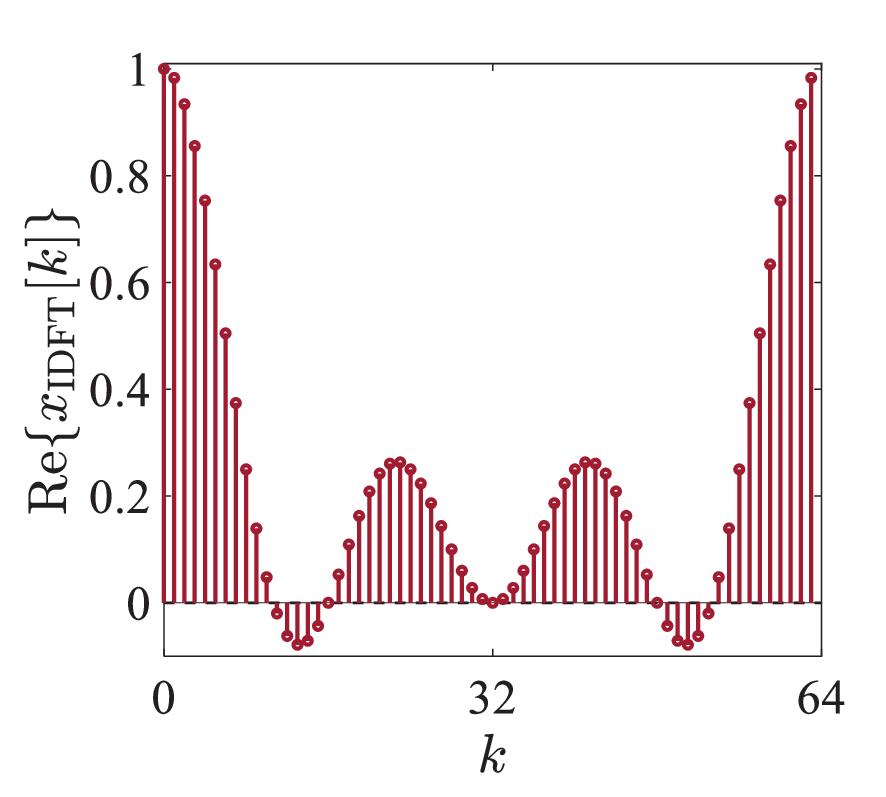}
        \captionsetup{skip=1pt}  
        \caption{}
        \label{fig:transmitted_symbols}
    \end{subfigure}
    \hfill
    \begin{subfigure}[t]{0.48\linewidth}
        \includegraphics[width=\linewidth]{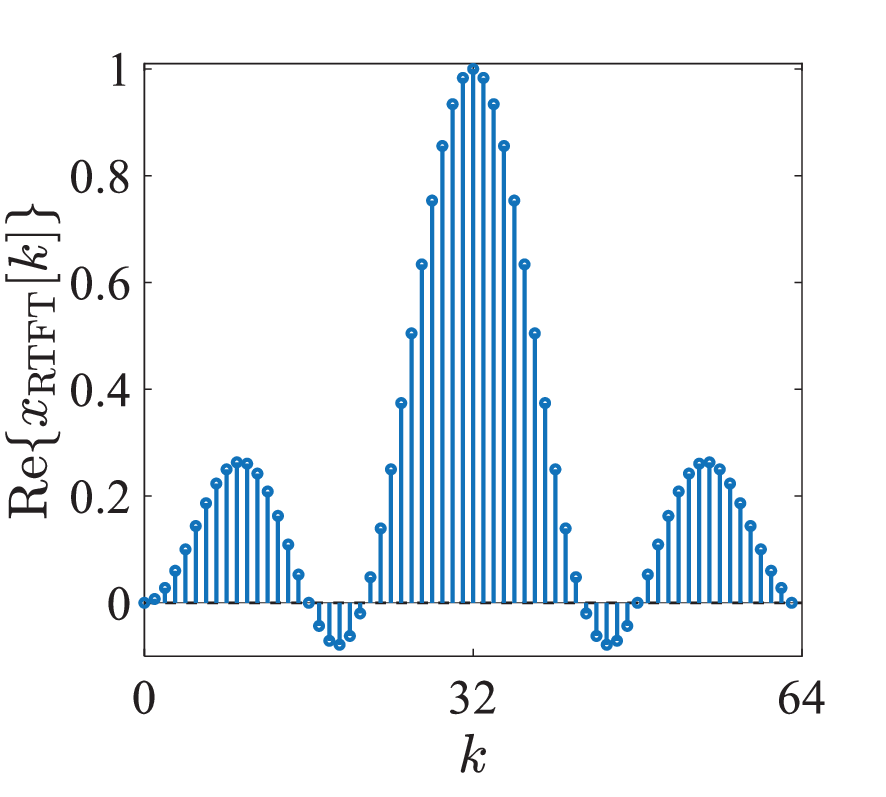}
        \captionsetup{skip=1pt}  
        \caption{}
        \label{fig:IFFT_symbols}
    \end{subfigure}

    \vspace{1mm} 
    \begin{subfigure}[t]{0.48\linewidth}
        \includegraphics[width=\linewidth]{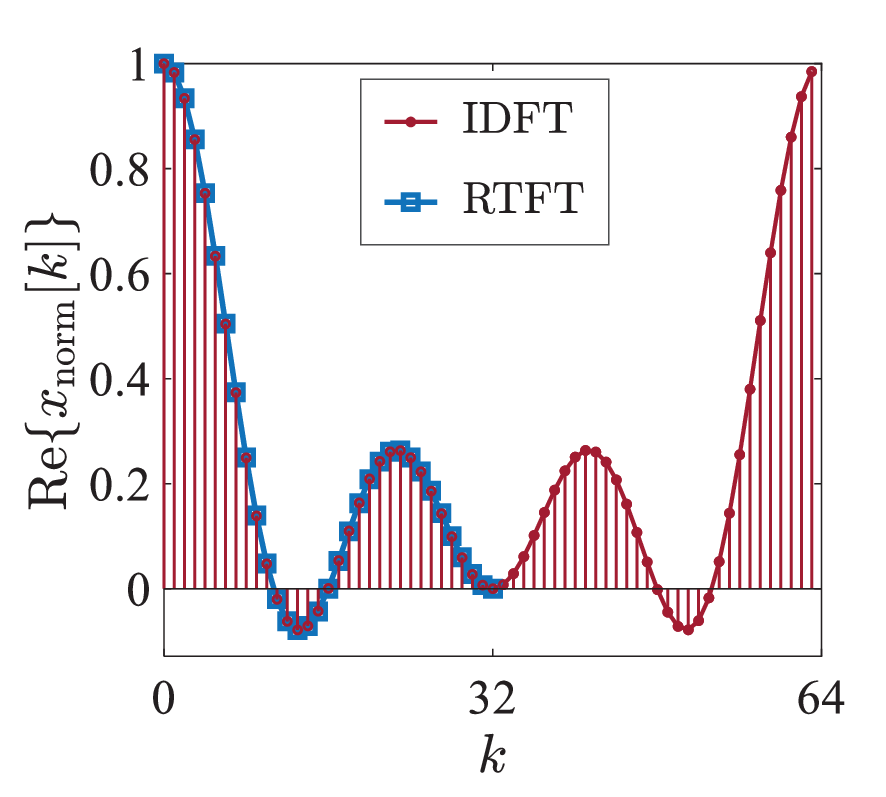}
        \captionsetup{skip=1pt}  
        \caption{}
        \label{fig:RTFT_OFDM_left}
    \end{subfigure}
    \hfill
    \begin{subfigure}[t]{0.48\linewidth}
        \includegraphics[width=\linewidth]{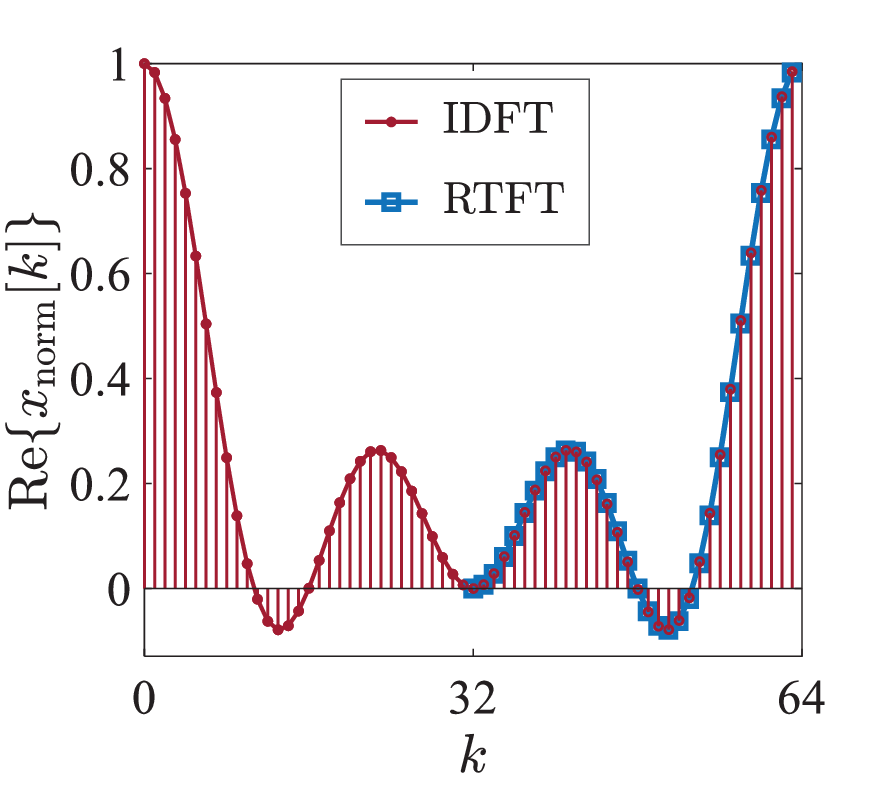}
        \captionsetup{skip=1pt}  
        \caption{}
        \label{fig:RTFT_OFDM_right}
    \end{subfigure}

    \caption{Comparison between IDFT-based [Eq.~\eqref{eq:OFDM_disc}] and RTFT-based [Eq.~\eqref{eq:OFDM_RTFT} or Fig.~\ref{fig:OFDM_RTFT}(b)] OFDM signals. 
    (a)~Normalized OFDM symbols (real part), $\text{Re}\{x_\text{IDFT}[k]\}$, obtained by applying IDFT to $\{X[n]\}$. 
    (b)~Normalized OFDM symbols (real part), $\text{Re}\{x_\text{RTFT}[k]\}$, obtained by sampling the RTFT output in Eq.~\eqref{eq:RTFT_OFDM_orth_rep_3}. 
    (c)~Comparison of the normalized IDFT-OFDM signal (real part) with the right part [see (b)] of the normalized RTFT-OFDM signal (real part). (d)~Comparison of the normalized IDFT-OFDM signal (real part) with the left part [see (b)] of the normalized RTFT-OFDM signal (real part).}
    \label{fig:compare_RTFT_IFFT}
\end{figure}

Figure~\ref{fig:compare_RTFT_IFFT} compares the conventional OFDM signal using IDFT and the proposed approach based on RTFT for the symbol sequence $\{X[n]\}$ in Fig.~\ref{fig:OFDM_RTFT}(a). Figure~\ref{fig:compare_RTFT_IFFT}(a) shows the real part of the normalized\footnote{$x_\text{norm}(t)=\frac{\text{Re}\{x(t)\}}{\max\{\text{Re}\{x(t)\}\}} + j\frac{\text{Im}\{x(t)\}}{\max\{\text{Im}\{x(t)\}\}}$.} OFDM symbols, $\text{Re}\{x_\text{IDFT}[k]\}$, obtained by applying IDFT to $\{X[n]\}$, as described in Eq.~\eqref{eq:OFDM_disc}. Figure~\ref{fig:compare_RTFT_IFFT}(b), on the other hand, presents the real part of the normalized OFDM symbols corresponding to the sampled output of the RTFT, as given in Eq.~\eqref{eq:RTFT_OFDM_disc}. Figures~\ref{fig:compare_RTFT_IFFT}(c) and~\ref{fig:compare_RTFT_IFFT}(d) further compare the IDFT-based signal with the left and right halves, respectively, of the RTFT-based signal. 

We note that the two (IDFT and RTFT) waveforms in Figs.~\ref{fig:compare_RTFT_IFFT}(a) and~\ref{fig:compare_RTFT_IFFT}(b) are shifted with respect to each other by the amount of $N/2$. This shift stems from the fact that the IDFT- OFDM and RTFT-OFDM systems map different periods of the input signal's periodic spectrum onto the time domain. Specifically, the IDFT-OFDM system operates over the interval $f_\text{in} \in \left[0, {2}/{T_\text{s}}\right]$, whereas the RTFT-OFDM system uses the interval $f_\text{in} \in \left[-{1}/{2T_\text{s}}, {1}/{2T_\text{s}}\right]$. However, as already announced, this difference has not consequence in the overall transceiver system.

\section{Performance Demonstration}
\label{sec:Demonstration}

\subsection{Accurate Recovery}
\label{sec:Demonstration_A}
\pati {Accurate Transmission and Recovery}

The transmitted symbols $\{X[n]\}$ can be recovered at the receiver through a second RTFT operation. To demonstrate that this can be accomplished accurately, we begin by considering an ideal scenario, where the adverse effects of the communication channel are neglected. Specifically, we assume the ideal transmission condition $y(t) = x(t)$ in Fig.~\ref{fig:RTFT-OFDM_system}, viz.,
\begin{equation} \label{eq:received_y_a}
\begin{aligned}
y(t) &= x_\text{RTFT}(t) \\[6pt]
     &= \sqrt{\frac{2\pi}{\left|\phi_{\text{tr},2}\right|}} 
        \int_{-\infty}^{\infty} X(\tau) 
        e^{j\frac{\phi_{\text{tr},1}+t}{\phi_{\text{tr},2}}\tau} d\tau,
\end{aligned}
\end{equation}
where $\phi_{\text{tr},2}$ and $\phi_{\text{tr},1}$ represent the phaser parameters at the transmitter side and are given by Eq.~\eqref{eq:phaser_cond1_phi2} and Eq.~\eqref{eq:phi1_value1} or Eq.~\eqref{eq:phi1_value2} respectively, viz.,
\begin{subequations} \label{eq:phaser_transmitter}
\begin{align}
\phi_{\text{tr},2} &= \pm\frac{T_0T_\text{s}}{2\pi} \label{eq:transmitter_phi2}, \\
\phi_{\text{tr},1} &= -T_0~\text{or} -{T_0}/{2}\label{eq:transmitter_phi1},
\end{align}
\end{subequations}
and where the expression for $x_\text{RTFT}(t)$ comes from Eq.~\eqref{eq:OFDM_RTFT}. The impact of the wireless channel will be addressed in the next subsection.

Next, the signal $y(t)$ is passed through an RTFT block at the receiver side. According to Eq.~\eqref{eq:RTFT_GVD}, the output $Y(t)$ is given by 
\begin{subequations} \label{eq:received_Y}
\begin{equation} \label{eq:received_Y_a}
\begin{aligned}
Y(t) = \sqrt{\frac{2\pi}{\left|\phi_{\text{re},2}\right|}} 
        \int_{-\infty}^{\infty} y(\tau^\prime) 
        e^{j\frac{\phi_{\text{re},1}+t}{\phi_{\text{re},2}}\tau^\prime} d\tau^\prime,
\end{aligned}
\end{equation}
where $\phi_{\text{re},2}$ and $\phi_{\text{re},1}$ are the phaser parameters at the receiver side. Substituting Eq.~\eqref{eq:received_y_a} into this equation yields
\begin{equation} \label{eq:received_Y_b}
\begin{aligned}
Y(t) = &\sqrt{\frac{2\pi}{\left|\phi_{\text{re},2}\right|}}\sqrt{\frac{2\pi}{\left|\phi_{\text{tr},2}\right|}}  
        \\
        & \int_{-\infty}^{\infty} \left[
        \int_{-\infty}^{\infty} X(\tau) 
        e^{j\frac{\phi_{\text{tr},1}+\tau^\prime}{\phi_{\text{tr},2}}\tau} d\tau\right] 
        e^{j\frac{\phi_{\text{re},1}+t}{\phi_{\text{re},2}}\tau^\prime} d\tau^\prime,
\end{aligned}
\end{equation}
which, upon exchanging the order of the integrals, becomes
\begin{equation} \label{eq:received_Y_c}
\begin{aligned}
Y(t) = &\sqrt{\frac{2\pi}{\left|\phi_{\text{re},2}\right|}}\sqrt{\frac{2\pi}{\left|\phi_{\text{tr},2}\right|}}\\
& \int_{-\infty}^{\infty} X(\tau)e^{j\frac{\phi_{\text{tr},1}}{\phi_{\text{tr},2}}\tau} \left[
        \int_{-\infty}^{\infty}  
        e^{j\left(\frac{\tau}{\phi_{\text{tr},2}}+\frac{\phi_{\text{re},1}+t}{\phi_{\text{re},2}}\right)\tau^\prime} d\tau^\prime\right] 
          d\tau
\end{aligned}
\end{equation}
and simplifies to
\begin{equation} \label{eq:received_Y_d}
\begin{aligned}
Y(t) = &2\pi\sqrt{\frac{2\pi}{\left|\phi_{\text{re},2}\right|}}\sqrt{\frac{2\pi}{\left|\phi_{\text{tr},2}\right|}}\\[2pt]
&\int_{-\infty}^{\infty} X(\tau)e^{j\frac{\phi_{\text{tr},1}}{\phi_{\text{tr},2}}\tau} \delta\left(\frac{\tau}{\phi_{\text{tr},2}}+\frac{\phi_{\text{re},1}+t}{\phi_{\text{re},2}}\right)d\tau
\end{aligned}
\end{equation}
which further simplifies to
\begin{equation} \label{eq:received_Y_e}
\begin{aligned}
Y(t) = &2\pi\sqrt{\frac{2\pi}{\left|\phi_{\text{re},2}\right|}}\sqrt{\frac{2\pi}{\left|\phi_{\text{tr},2}\right|}}\\[2pt]
& \phi_{\text{tr},2} X\left(-\phi_{\text{tr},2}\frac{\phi_{\text{re},1} + t}{\phi_{\text{re},2}} \right) e^{-j \left( \phi_{\text{tr},1} \frac{\phi_{\text{re},1} + t}{\phi_{\text{re},2}} \right)} .
\end{aligned}
\end{equation}
\end{subequations}

Inspecting this relation reveals that proper signal recovery, $Y(t)=X(t)$, implies the twofold condition
\begin{subequations} \label{eq:phaser_receiver}
\begin{align}
\phi_{\text{re},1} &=0 \label{eq:receiver_phi1}, \\
\phi_{\text{re},2} &= - \phi_{\text{tr},2}\label{eq:receiver_phi2},
\end{align}
\end{subequations}
with the latter conditions representing \emph{inverse} group delay dispersion, making the receiver Fourier transformation an RT-FT forming a pair with the transmitter RT-IFT operation. The practical implementability of parameter values in Eqs.~\eqref{eq:phaser_transmitter} and~\eqref{eq:phaser_receiver} is discussed in Appendix~\ref{sec:appendix_E_feasibility}.

Substituting Eqs.~\eqref{eq:receiver_phi1} and~\eqref{eq:receiver_phi2} into Eq.~\eqref{eq:received_Y_e}
\begin{subequations} \label{received_Y_final}
\begin{equation} \label{eq:received_Y_final_a}
Y(t) =4\pi^2X(t)e^{j\frac{\phi_{\text{tr},1}}{\phi_{\text{tr},2}}t},
\end{equation}
and further substituting $t=nT_\text{s}$ and Eq.~\eqref{eq:transmitter_phi2} into this relation yields
\begin{equation} \label{eq:received_Y_final_b}
Y(nT_\text{s}) =4\pi^2X(nT_\text{s})e^{\pm j 2\pi n \frac{\phi_{\text{tr},1}}{T_0}}.
\end{equation}
\end{subequations}

\begin{figure}[h!]
    \centering
    \includegraphics[width=0.8\linewidth]{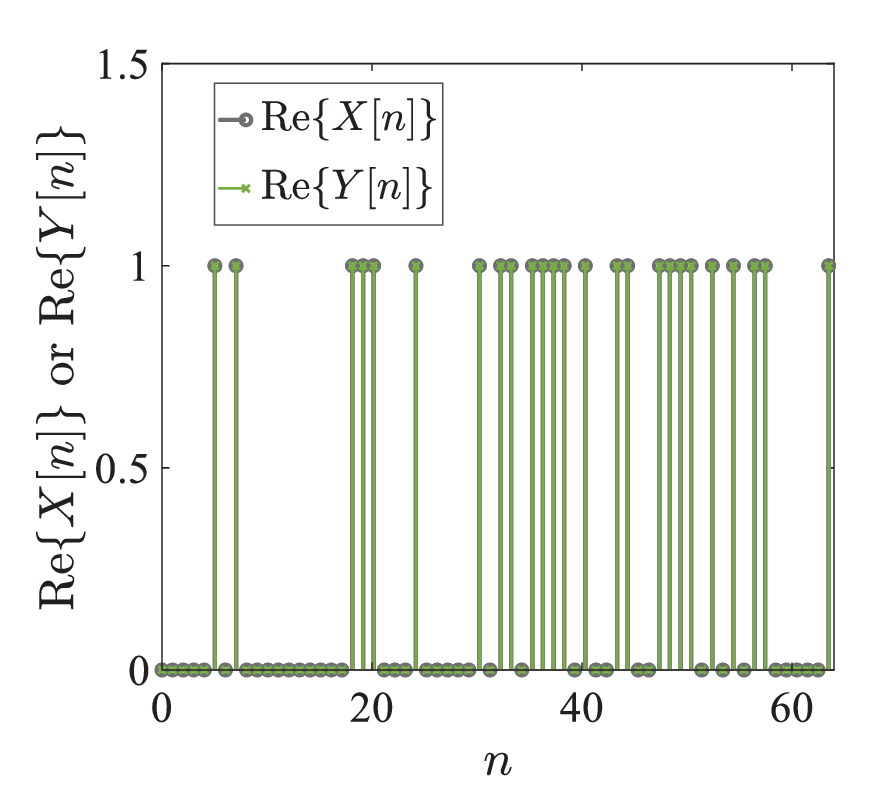}
    \caption{Data recovery (real parts) by the RTFT-based OFDM system (Fig.~\ref{fig:RTFT-OFDM_system}) for a random sequence $\{X[n]\}$ of $N=64$ symbols.} \label{fig:TR_symbols}
\end{figure}

As in the transmitter case, the parameter $\phi_{\text{tr},1}$ can take the values $-T_0/2$ and $-T_0$. Substituting these two choices into Eq.~\eqref{eq:received_Y_final_b} leads to
\begin{subequations} \label{received_Y_differ_phi1}
\begin{align}
Y[n]&=X[n]e^{\pm j n\pi}\label{eq:received_Y_phi1_1} \\
&\text{and}\nonumber  \\
Y[n]&=X[n],\label{eq:received_Y_phi1_2}
\end{align}
\end{subequations}
respectively, where the latter solution readily provides recovery, while the former solution provides recovery after straightforward multiplication by $e^{\mp j n\pi}=(-1)^n$. An example of successful recovery for a random data sequence is provided in Fig.~\ref{fig:TR_symbols}.

\subsection{Robustness to Multipath Fading}
\label{sec:Demonstration_B}
\pati{Robustness to multipath Fading}
Section~\ref{sec:Demonstration_A} established that, in the absence of channel impairments, the proposed analog OFDM system, with appropriately designed phaser parameters, enables accurate transmission and recovery of data symbols. In this section, we extend the analysis to realistic wireless environments and evaluate the robustness of the proposed system against multipath fading.

Since multipath fading generally occurs between different data symbols~\cite{goldsmith2005wireless}, we write the overall baseband OFDM signal as
\begin{subequations}
    \begin{equation}\label{eq:signal_mblocks}
        x(t)=\sum_{m}x_m(t),
    \end{equation}
    where the $m^\textrm{th}$ signal block reads [Eq.~\eqref{eq:OFDM_signal_a_timedomain}]
    \begin{equation}\label{eq:block_m}
        x_m(t)=\sum_{n=0}^{N-1}X_m[n]w_m(t)\varphi_n(t),
    \end{equation}
    with
    \begin{equation}\label{eq:window_m}
        w_m(t)=\Pi\left(\frac{t-(m+1)T_0/2}{T_0}\right).
    \end{equation}
\end{subequations}
Assuming a linear time-invariant (LTI) channel without noise, the received baseband signal is given by
\begin{equation} \label{eq:y_m_RTFT_1}
\begin{aligned}
y_m(t)=x_m(t)*h(t),
\end{aligned}
\end{equation}
where $h(t)$ denotes the channel impulse response. In a typical multipath fading environment, this response can be modeled as~\cite{biglieri2002fading}
\begin{equation} \label{eq:channel_response}
\begin{aligned}
h(t)=\sum_{l=0}^{L-1}\gamma_l\delta(t-\tau_l),
\end{aligned}
\end{equation}
where $L$ is the number of the multiple propagation paths, and $\gamma_l$ and $\tau_l$ denote the complex gain and time delay of the $l^{\text{th}}$ path, respectively.

Assuming that the line-of-sight (LoS) component is dominant, then $\gamma_l$ is generally modeled as a non-zero-mean complex Gaussian variable, whose magnitude has a Rice probability distribution and whose phase is uniformly distributed in the interval $(0,2\pi)$~\cite{proakis2008digital}. The probability density function (PDF) of $\left|\gamma_l\right|=R$ is thus\footnote{Uppercase $R$ represents the random variable itself, whose exact value is unknown until a random experiment is conducted. Lowercase $r$ denotes a specific realization (value) of that random variable.}
\begin{equation} \label{eq:Rice_PDF}
\begin{aligned}
p_{R}(r)=\frac{r}{\sigma^2}e^{-\frac{r^2+s^2}{2\sigma^2}}I_0\left(\frac{rs}{\sigma^2}\right),~~~~r\geq0,
\end{aligned}
\end{equation}
where $s=|\mu|$ denotes the amplitude\footnote{When $s=0$, Eq.~\eqref{eq:Rice_PDF} reduces to the Rayleigh distribution $p_{R}(r)=\frac{r}{\sigma^2}e^{-\frac{r^2}{2\sigma^2}},~r\geq0$, with $\sigma^2=E(R^2)/2$.} of the deterministic LoS component ($|\gamma_0|$) and $\sigma^2$ represents the variance of the scattered (Gaussian) components. $I_0(\cdot)$ is the zero-order modified Bessel function of the first kind.

Substituting Eq.~\eqref{eq:channel_response} into Eq.~\eqref{eq:y_m_RTFT_1}, the received signal for the $m^{\text{th}}$ OFDM block becomes
\begin{equation} \label{eq:y_m_RTFT_2}
\begin{aligned}
y_m(t)=\sum_{l=0}^{L-1}\gamma_l x_m(t-\tau_l),
\end{aligned}
\end{equation}
where $\gamma_l$ and $\tau_l$ have the statistical properties described in the previous paragraph.

We now need to discretize Eq.~\eqref{eq:y_m_RTFT_2}. Assuming uniformly spaced path delays, such that $\tau_l = \tau_0 + lT_\text{s}$, and evaluating the received signal at the sampling instants $t = \tau_0 + nT_\text{s}$, we obtain for the discrete-time form of the received signal as
\begin{equation} \label{eq:y_m_RTFT_dis}
\begin{aligned}
y_m[n]=\sum_{l=0}^{L-1}h[l] x_m\left[n-l\right],~~~0\leq n \leq N-1,
\end{aligned}
\end{equation}
where $x_m[n]$ and $y_m[n]$ denote the $n^{\text{th}}$ transmitted and received symbols, respectively, of the $m^{\text{th}}$ OFDM block. As apparent in Eq.~\eqref{eq:y_m_RTFT_dis}, the received symbols, $y_m[n]$, are composed of delayed versions of the transmitted OFDM symbols, $x_m[n-l]$, weighted by the corresponding complex channel gain, $h[l]$. In other words, each received symbol is affected by previous symbols due to multipath delays, leading to persistent ISI.

To eliminate the ISI introduced by the convolution in Eq.~\eqref{eq:y_m_RTFT_dis}, a prefix of length of $L$ (or more) can be added to the beginning of each OFDM block. This prefix ensures that the linear convolution does not interfere with the data-bearing portion of the OFDM symbol. A detailed explanation of this technique is provided Appendix~\ref{sec:appendix_B_prefix}. Commonly used prefixes include cyclic prefix (CP) or zero-padding prefix (ZP). 

Let us consider CP, which is usually preferred, due to its more efficient---one-tap DFT-based equalization---transformation of linear convolution into circular convolution. Inserting a CP of length $L$ transforms the transmitted symbols in the $m^\text{th}$ OFDM block, indexed by $-L \leq n \leq N-1$, to $\{x_m^{\text{cp}}[n]\} = \{x_m[N-L], \dots, x_m[N-1], x_m[0], \dots, x_m[N-1]\}$. The corresponding received symbols, as expressed in Eq.~\eqref{eq:y_m_RTFT_dis}, are then given by
\begin{equation} \label{eq:y_m_RTFT_zp}
\begin{aligned}
y^{\text{cp}}_{m}[n]=\sum_{l=0}^{L-1}h[l] x^{\text{cp}}_{m}[n-l],~~-L\leq n\leq N-1.
\end{aligned}
\end{equation}
Discarding the first $L$ samples of this signal leaves out only the symbols affected by the transmitted symbols within the same OFDM block. Denoting this portion $\{y^\prime_{m}[n]\}$, we have
\begin{equation} \label{eq:y_m_RTFT_remove_zp}
\begin{aligned}
y^\prime_{m}[n]=\sum_{l=0}^{L-1}h[l] x_{m}[n-l]_N,~~0\leq n\leq N-1,
\end{aligned}
\end{equation}
which implies that these symbols form a circular convolution of the transmitted sequence $x_m[n]$ and the channel impulse response $h[l]$, i.e.,
\begin{equation} \label{eq:y_m_RTFT_circular_conv}
\begin{aligned}
y^\prime_{m}[n] = h[l] \circledast x_m[n],\quad 0 \leq n \leq N - 1,
\end{aligned}
\end{equation}
where $\circledast$ denotes the circular convolution product. 

In the discrete-time domain, circular convolution in the time domain corresponds to element-wise multiplication in the frequency domain. Applying the RTFT—using the phaser parameters specified in Eq.~\eqref{eq:phaser_receiver}—to both sides of Eq.~\eqref{eq:y_m_RTFT_circular_conv}, we obtain then
\begin{equation} \label{eq:Ym_RTFT}
\begin{aligned}
Y^\prime_m[i]&=\text{RTFT}\{y^\prime_m[n]\}\\
&=\text{RTFT}\{{h}[n] \circledast x_m[n]\}\\
&={H}[i]X_m[i],\quad 0 \leq i \leq N - 1,
\end{aligned}
\end{equation}
which shows that the transmitted symbols $\{X_m[i]\}$ can be exactly recovered via the trivial operation $X_m[i]=Y^\prime_m[i]/{H}[i]$, where $\{{H}[i]\}$ is the RTFT of the channel response ${h}[l]$.

\pati{BER Performance }
To further assess the robustness of the proposed RTFT-based OFDM system under realistic transmission conditions, we extend the analysis to account for additive noise. In this case, transmission is considered to occur across a multipath Rician fading channel with additive white Gaussian noise (AWGN). The performance of the proposed RTFT-based OFDM system is quantitatively compared against that of the conventional IFFT-based OFDM system in terms of bit error rate (BER) and symbol constellation distribution.

The BER evaluation is conducted via a Monte Carlo simulation\footnote{In the Monte Carlo approach, the BER at each $E_b/N_0$ point is obtained by averaging over multiple independent realizations of the transmission process, in this case $1000$ OFDM blocks per point.} across the SNR range $E_b/N_0 = 0$–30~dB, where $E_b$ denotes the energy per bit and $N_0$ represents the one-sided power spectral density of the AWGN. The Rician channel model consists of four NLoS paths. Both RTFT-OFDM and IFFT-OFDM systems employ $N=64$ subcarriers with QPSK modulation, and a prefix of length $L=5$ is used in the simulations. 

\begin{figure}[h!]
    \centering
    \begin{subfigure}[t]{0.48\linewidth}
        \includegraphics[width=\linewidth]{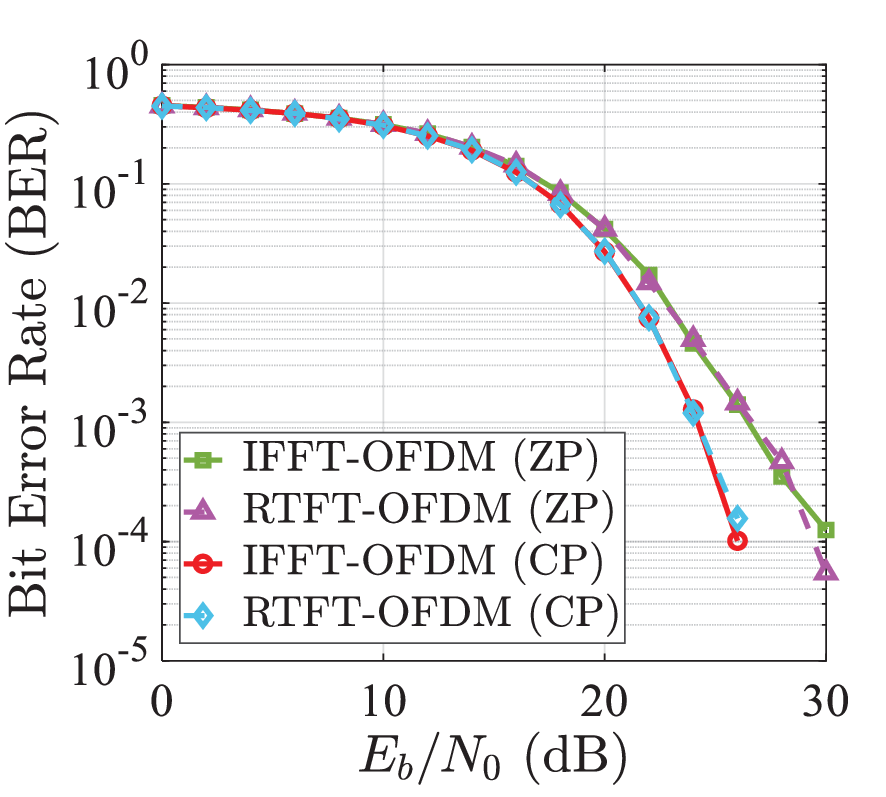}
        \captionsetup{skip=1pt}  
        \caption{}
        \label{fig:BER}
    \end{subfigure}
    \hfill
    \begin{subfigure}[t]{0.48\linewidth}
        \includegraphics[width=\linewidth]{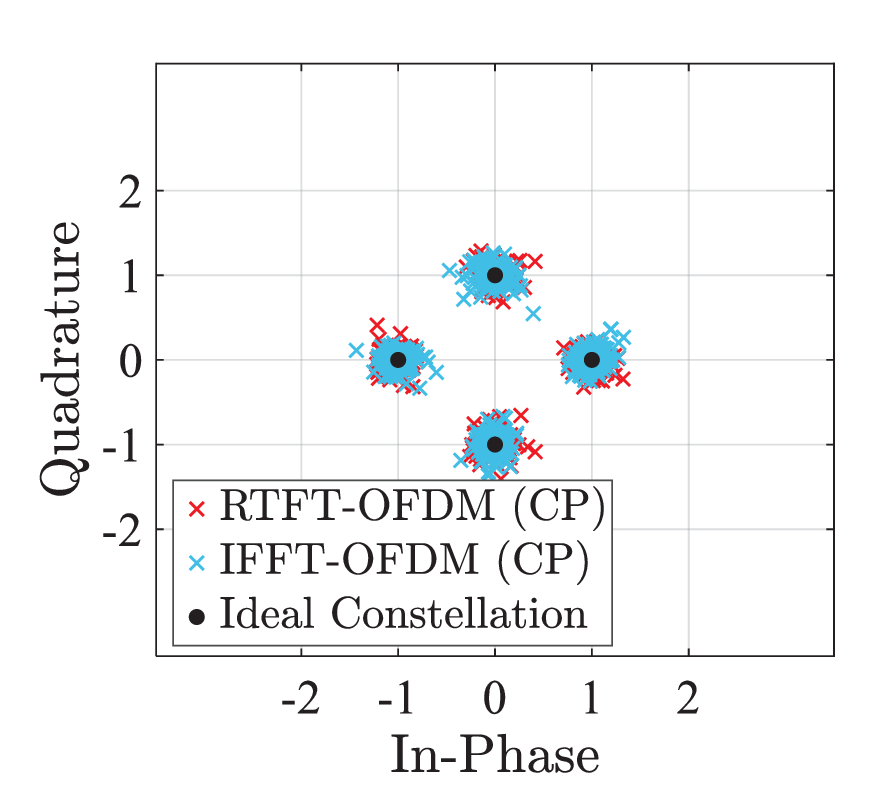}
        \captionsetup{skip=1pt}  
        \caption{}
        \label{fig:Constellation}
    \end{subfigure}
    \caption{Performance comparison between RTFT-OFDM and IFFT-OFDM systems with $N=64$ subcarriers and QPSK modulation over a Rician fading channel with AWGN. (a)~BER performance for both CP and ZP schemes, obtained via Monte Carlo simulation over $1000$ OFDM blocks for each SNR point. (b)~Received symbol constellation diagram (CP case) at $E_b/N_0=30$~dB.}
    \label{fig:BER_performance}
\end{figure}

The simulation results are presented in Fig.~\ref{fig:BER_performance}. ZP results are also shown for comparison, as ZP is more amenable to implementation in the proposed analog system using microwave delay-line technology\footnote{In fact, despite its linear convolution nature, ZP becomes equivalent to CP once the first $L$ received samples are discarded and the channel is zero-padded at the receiver.}. Figure~\ref{fig:BER_performance}(a) shows that the RTFT-based OFDM achieves nearly identical BER performance to the conventional IFFT-based OFDM in both CP and ZP schemes. This confirms that the proposed analog RTFT system is capable of achieving robust transmission in multipath environments in the presence of noise. It is also observed that the ZP configuration exhibits slightly inferior BER performance compared to the CP case, particularly at high SNR. This behavior arises from fundamental differences in noise handling: while CP converts linear convolution into a perfect circular convolution, preserving the white noise characteristics across subcarriers, ZP requires a numerical reconstruction of circularity, which slightly colors and amplifies the noise. At low SNR, this effect is negligible compared to the dominant noise floor, leading to nearly identical BER. At high SNR, however, the relative impact of the noise enhancement becomes more pronounced, resulting in a small BER penalty for ZP. Figure.~7(b) shows the corresponding received symbol constellations for the CP case with QPSK at $E_b/N_0 = 30$~dB. The received symbols for both systems are tightly clustered around the ideal QPSK constellation points, indicating negligible distortion. This result further confirms that the analog RTFT implementation preserves signal integrity at high SNR.

\section{Conclusion}
\label{sec:Conclusion}
\pati {Advantages}
An analog OFDM system based on RTFT has been proposed, theoretically analyzed and numerically validated. Replacing digital FFT/IFFT processors with passive microwave phaser circuits engineered for appropriate GVD, the proposed system enables real-time frequency-to-time mapping in the analog domain. Simulation results confirm the theoretical equivalence between RTFT-based analog OFDM and conventional OFDM, and demonstrate the system’s robustness to multipath fading when prefixing is applied. Owing to its ultra-fast operation, low power consumption, and elimination of digital complexity, this architecture shows strong potential for future wideband, high-speed wireless communication systems, particularly in millimeter-wave, terahertz, and optical frequency regimes. Future research will focus on overcoming the limitations associated with realizing large group delay dispersion. These efforts aim to enhance the scalability and broaden the practical applicability of analog OFDM systems across a wider range of frequency bands and data rates.

\begin{appendices}
\section{Benefit of the Narrow-band subcarriers in OFDM}
\label{sec:appendix_A_narrowband}

To better understand how the narrow-band subcarriers in OFDM scheme contribute to reducing the ISI, we compare the modulated signal on the $n^{\text{th}}$ subcarrier in OFDM scheme with the modulated signal in a wideband single-carrier (WSC) modulation scheme.

\pati{Narrow-band subcarriers for Reducing ISI}

According to Eq.~\eqref{eq:block_m}, the modulated signal on the $n^{\text{th}}$ subcarrier in the $m^{\text{th}}$ OFDM block is expressed as
\begin{subequations}\label{eq:nth_carrier_oneblock}
\begin{equation}\label{eq:nth_carrier_a}
c_{n,m}(t)=X_m[n]w_m(t)\varphi_n(t),
\end{equation}
where $X_m[n]$ denotes the $n^{\text{th}}$ data symbol in the $m^{\text{th}}$ block, $w_m(t)$ is the rectangular window function applied on the $m^{\text{th}}$ block, which is given by Eq.~\eqref{eq:window_m}, and $\varphi_n(t)$ is the $n^{\text{th}}$ subcarrier waveform, given by Eq.~\eqref{eq:OFDM_signal_c_orth_sub}. Substituting then Eq.~\eqref{eq:window_m} and Eq.~\eqref{eq:OFDM_signal_c_orth_sub} into Eq.~\eqref{eq:nth_carrier_a} leads to
\begin{equation}\label{eq:nth_carrier}
c_{n,m}(t)=A_n\Pi\left(\frac{t-(m+1)T_0/2}{T_0}\right)X_m[n]e^{j2\pi f_n t},
\end{equation}
\end{subequations}
which leads to the following modulated signal along the whole time domain on the $n^{\text{th}}$ subcarrier:
\begin{equation}\label{eq:nth_carrier_alltime}
r_n(t)=\sum_{m}c_{n,m}(t),
\end{equation}
where $n=0,1,\dots,N-1$, $m \in \mathbb{Z}_{+}$.

\begin{figure}[!h]
    \centering
    \begin{subfigure}[t]{0.55\linewidth}
        \includegraphics[width=\linewidth]{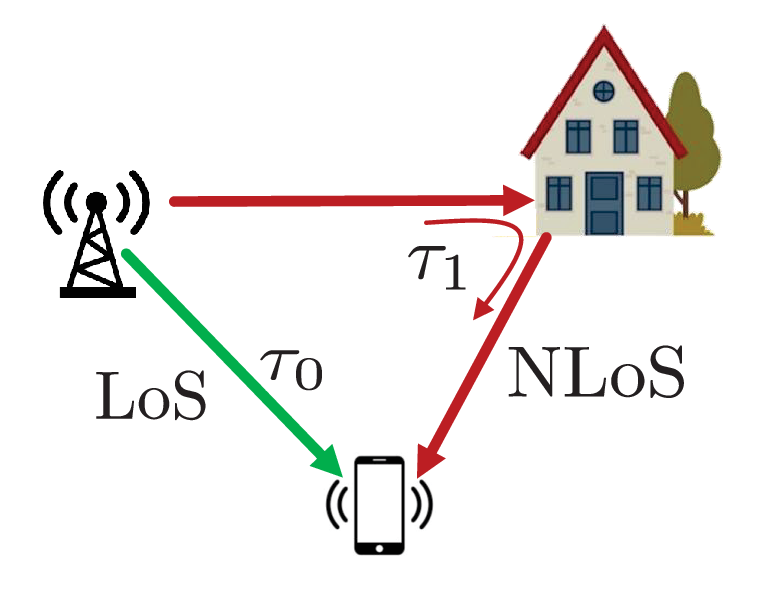}
        \captionsetup{skip=1pt}  
        \caption{}
        \label{fig:multipath_a}
    \end{subfigure}
    \vspace{0.1cm} 
    \begin{subfigure}[t]{0.9\linewidth}
        \includegraphics[width=\linewidth]{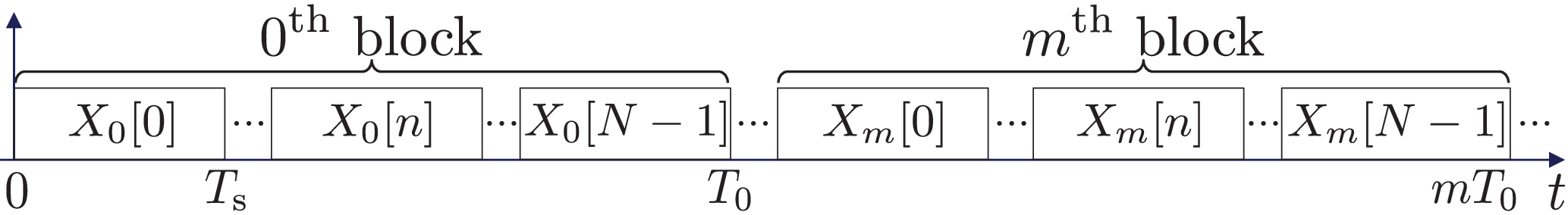}
        \captionsetup{skip=1pt}  
        \caption{}
        \label{fig:block_fs_b}
    \end{subfigure}
    \vspace{0.1cm} 
    \begin{subfigure}[t]{0.9\linewidth}
        \includegraphics[width=\linewidth]{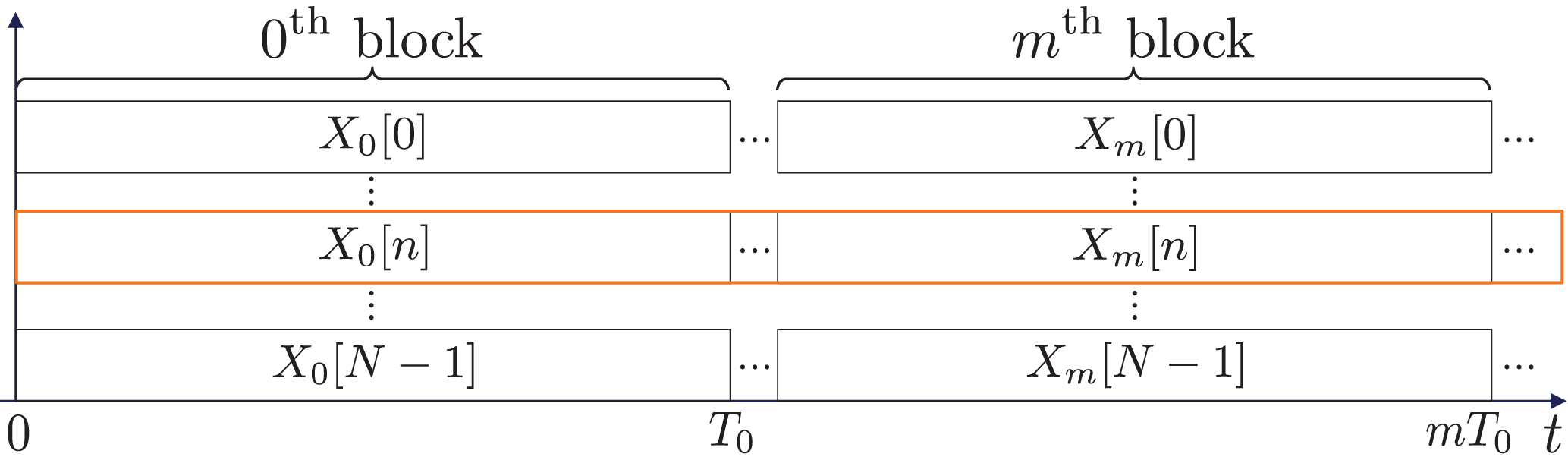}
        \captionsetup{skip=1pt}  
        \caption{}
        \label{fig:block_fn_c}
    \end{subfigure}
    \vspace{0.1cm} 
    \begin{subfigure}[t]{0.48\linewidth}
        \includegraphics[width=\linewidth]{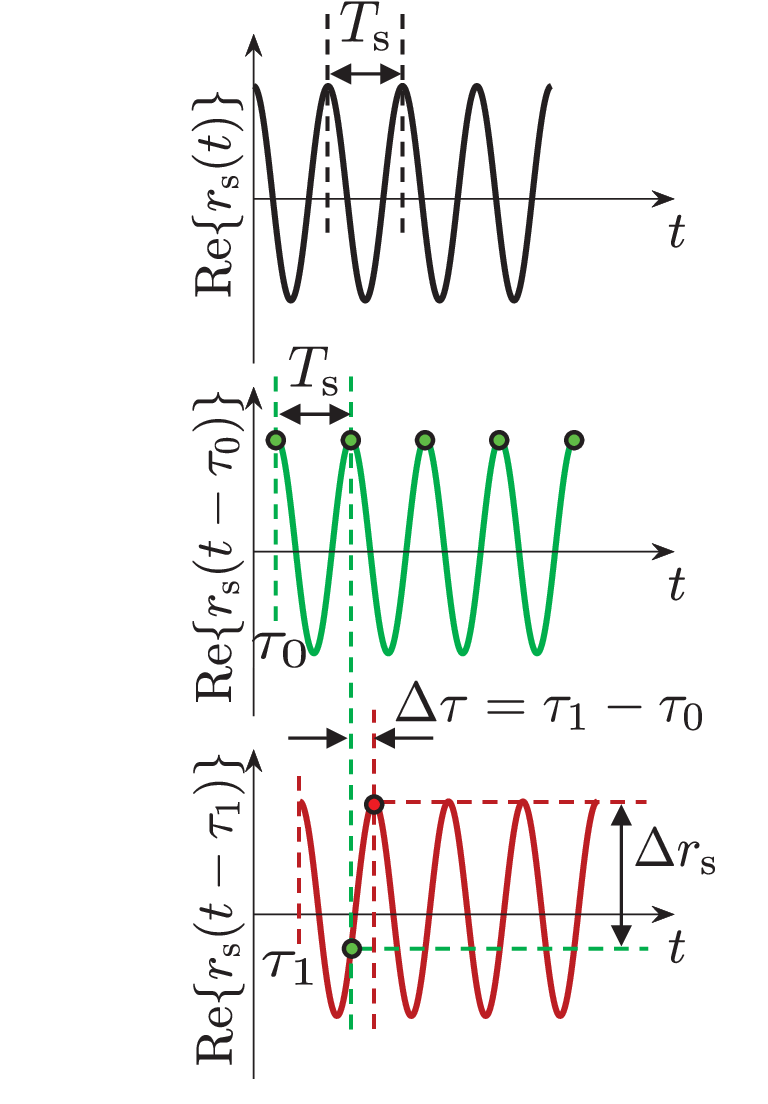}
        \captionsetup{skip=1pt}  
        \caption{}
        \label{fig:ISI_fs_d}
    \end{subfigure}
    \hspace{0.01cm}
    \begin{subfigure}[t]{0.48\linewidth}
        \includegraphics[width=\linewidth]{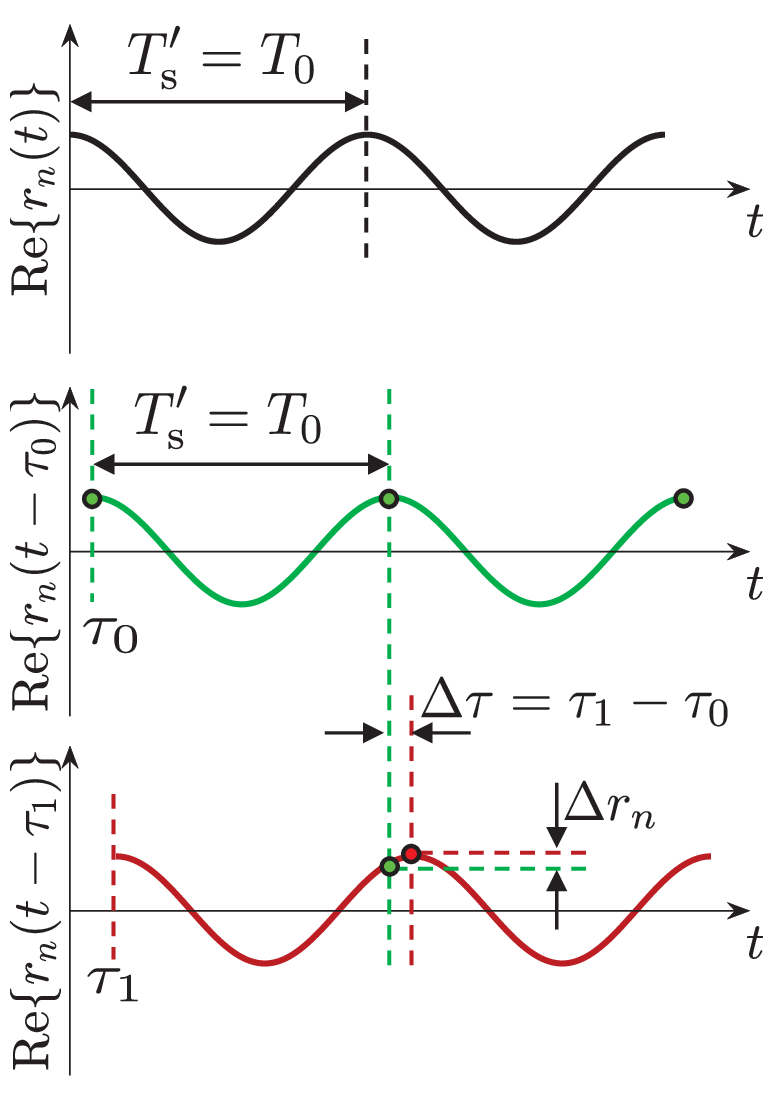}
        \captionsetup{skip=1pt}  
        \caption{}
        \label{fig:ISI_fn_e}
    \end{subfigure}

    \caption{Comparison of inter-symbol interference (ISI) in a multipath environment for two different modulation schemes. 
    (a)~Typical multipath scenario, featuring a line-of-sight (LoS) path, with delay $\tau_0$, and a non-line-of-sight (NLoS) path, with delay $\tau_1$. 
    (b)~Time domain data stream for the transmitted data symbols in a wideband single-carrier (WSC) modulation scheme. 
    (c)~Same as (b) but in the orthogonal-frequency division multiplexing (OFDM) modulation scheme, with highlighted block corresponding to the $n^{\text{th}}$ subcarrier. 
    (d)~Transmitted and received signals in the WSC scheme. Due to the short symbol duration, $T_\text{s}$, the delay difference $\Delta \tau=\tau_1-\tau_0$ introduces significant distortion (quantified by $\Delta r_\text{s}$), resulting in large ISI. 
    (e)~Same as (d) but in the OFDM scheme, for the subcarrier $\varphi_n(t)$. The much longer symbol period, $T_\text{s}'=T_0=NT_\text{s}$, allows the same delay difference $\Delta \tau$ to introduce less distortion (quantified by $\Delta r_n$), resulting in reduced ISI.}
    \label{fig:ISI}
\end{figure}

In the WSC modulation scheme, the modulated signal on the single-carrier $\varphi_{\text{s}}(t)$ in the $m^{\text{th}}$ OFDM block is 
\begin{subequations}\label{eq:single_carrier}
\begin{equation}\label{eq:single_carrier_a_time}
c_{\text{s},m}(t)=\sum_{n=0}^{N-1}X_m[n]w_{n,m}(t)\varphi_{\text{s}}(t),
\end{equation}
where $w_{n,m}(t)$ is the rectangular window function applied to the $n^{\text{th}}$ symbol in the $m^{\text{th}}$ block,
\begin{equation}\label{eq:single_carrier_b_window}
w_{n,m}(t)=\Pi\left(\frac{t-mT_0-(n+1)T_\text{s}/2}{T_\text{s}}\right),
\end{equation}
to be illustrated in Fig.~\ref{fig:ISI}(C), and where carrier waveform $\varphi_{\text{s}}(t)$ is given by  
\begin{equation}\label{eq:single_carrier_c_carrier}
\varphi_{\text{s}}(t)=e^{j2\pi f_\text{s} t},
\end{equation}
with the carrier frequency\footnote{In the WSC scheme, the carrier frequency is selected as $f_\text{s}=1/T_\text{s}$ to achieve the same data rate as in the OFDM scheme.}
\begin{equation}\label{eq:single_carrier_d_carrier_freq}
f_\text{s}=\frac{1}{T_\text{s}}.
\end{equation}
\end{subequations}
Substituting Eq.~\eqref{eq:single_carrier_b_window} and Eq.~\eqref{eq:single_carrier_c_carrier} into Eq.~\eqref{eq:single_carrier_a_time} yields
\begin{equation}\label{eq:single_carrier_time}
c_{\text{s},m}(t)=\sum_{n=0}^{N-1}\Pi\left(\frac{t-mT_0-(n+1)T_\text{s}/2}{T_\text{s}}\right)X_m[n]e^{j2\pi f_\text{s} t},
\end{equation}
which leads to the following modulated signal along the whole time domain on the single carrier $\varphi_{\text{s}}(t)$:
\begin{equation}\label{eq:nth_carrier_alltime}
r_\text{s}(t)=\sum_{m}c_{\text{s},m}(t).
\end{equation}

Comparing Eq.~\eqref{eq:single_carrier_time} and Eq.~\eqref{eq:nth_carrier} shows that the WSC scheme involves a short short symbol duration ($T_\textrm{s}$), whereas OFDM involves much longer symbol duration ($T_0=NT_\textrm{s}$) for each of its subcarriers.

Figure~\ref{fig:ISI} provides a visual comparison of the ISI behavior in both schemes under a typical multipath propagation scenario, which includes a line-of-sight (LoS) path with delay $\tau_0$ and a non-line-of-sight (NLoS) path with delay $\tau_1$, as shown in Fig.~\ref{fig:ISI}(a). Figure~\ref{fig:ISI}(b) illustrates the structure of the transmitted data stream in the WSC scheme, where each data symbol occupies the short time duration $T_\text{s}$, while Fig.~\ref{fig:ISI}(c) shows the same stream of the OFDM scheme, with significantly longer, $T_0$-long, symbols on each subcarrier. The impact of the corresponding multipath delays is illustrated in Fig.~\ref{fig:ISI}(d) and Fig.~\ref{fig:ISI}(e). In the WSC case (Fig.~\ref{fig:ISI}(d)), due to the short symbol duration ($T_\text{s}$), the delay difference $\Delta \tau=\tau_1-\tau_0$ introduces significant distortion (quantified by $\Delta r_\text{s}$), resulting in large ISI. In contrast, in the OFDM system (Fig.~\ref{fig:ISI}(e)), the longer symbol duration ($T_0$) allows the same delay difference $\Delta \tau$ to introduce much less distortion (quantified by $\Delta r_n$), resulting in reduced ISI.
\setlength{\abovecaptionskip}{0pt}

To support this observation, we provide next an analytical comparison of the ISI distortion experienced in the two systems. For the WSC modulation scheme, the ISI caused by multipath fading can be observed between two adjacent data symbols within a single OFDM block, e.g., between the $(n-1)^{\text{th}}$ and the $n^{\text{th}}$ symbols in the $m^{\text{th}}$ OFDM block. The detection clock\footnote{This assumes that the sampling clock is aligned with the arrival time of the LoS path, meaning that each symbol is sampled at the moment it first reaches the receiver via the LoS path.} for the $n^{\text{th}}$ symbol at the receiver is given by $t_n=\tau_0+nT_\text{s}$. When $n\geq 1$, the detection for the $n^{\text{th}}$ symbol will experience interference from the preceding symbol. This distortion can be quantified as the difference in the received signal amplitude at the sampling time $t_n$ between the two different paths, as illustrated in Fig.~\ref{fig:ISI}(d), $\Delta r_\text{s}$ can be expressed as
\begin{subequations} \label{eq:ISI_fs} 
\begin{equation}\label{eq:ISI_fs_a}
\begin{aligned}
\Delta r_\text{s}=\left|\text{Re}\left\{X_m[n]\varphi_\text{s}(t_n-\tau_0)\right\}-\text{Re}\left\{X_m[n-1]\varphi_\text{s}(t_n-\tau_1)\right\}\right|.
\end{aligned}
\end{equation}
Assuming $X_m[n-1]=X_m[n]=1$, using Eq.~\eqref{eq:single_carrier_c_carrier} and substituting $t_n=\tau_0+nT_\text{s}$, Eq.~\eqref{eq:ISI_fs_a} becomes
\begin{equation}\label{eq:ISI_fs_b}
\begin{aligned}
\Delta r_\text{s}&=\left|\cos{(2\pi n f_\text{s} T_\text{s})}-\cos{(2\pi f_\text{s} (nT_\text{s}-\Delta \tau))}\right|\\
&=\left|\cos{\left(2\pi n \frac{1}{T_\text{s}} T_\text{s}\right)}-\cos{\left(2\pi \frac{1}{T_\text{s}} (nT_\text{s}-\Delta \tau)\right)}\right|\\
&=\left|1-\cos(2\pi f_\text{s} \Delta \tau)\right|\\
&=2\sin^2{\left(\pi\frac{\Delta \tau}{T_\text{s}}\right)}.
\end{aligned}
\end{equation}
\end{subequations}
For the OFDM scheme, the ISI on the $n^{\text{th}}$ subcarrier can occurs between two adjacent blocks, e.g., between $X_{m-1}[n]$ and $X_{m}[n]$, and the detection clock for the $X_{m}[n]$ at the receiver is given by $t_m=\tau_0+mT_0$. When $m\geq 1$, the detection for the $X_{m}[n]$ symbol will experience interference from the preceding symbol. This distortion can be quantified as the difference in the received signal amplitude at the sampling time $t_m$ between the two different paths, as illustrated in Fig.~\ref{fig:ISI}(e), and the distortion is quantified as
\begin{subequations} \label{eq:ISI_fn} 
\begin{equation}\label{eq:ISI_fn_a}
\Delta r_n=\left|\text{Re}\left\{X_{m}[n]\varphi_n(t_m-\tau_0)\right\}-\text{Re}\left\{X_{m-1}[n]\varphi_n(t_m-\tau_1)\right\}\right|.
\end{equation}
Assuming again $X_m[n]=X_{m-1}[n]=1$, Eq.~\eqref{eq:ISI_fn_a} and Eq.~\eqref{eq:OFDM_signal_c_orth_sub} and $t_m=\tau_0+mT_0$ into this expression yields
\begin{equation}\label{eq:ISI_fn_b}
\begin{aligned}
\Delta r_{n}&=A_n\left|\cos{(2\pi m f_n T_0)}-\cos{(2\pi f_n (mT_0-\Delta \tau))}\right|\\
&= A_n\left|\cos{\left(2\pi m \frac{n}{T_0} T_0\right)}-\cos{\left(2\pi \frac{n}{T_0} (mT_0-\Delta \tau)\right)}\right| \\
&= A_n\left|1-\cos{\left(2\pi \frac{n}{T_0} \Delta \tau\right)}\right|\\
&=2A_n\sin^2{\left(\pi n\frac{\Delta \tau}{T_0}\right)}\\
&=2A_n\sin^2{\left(\pi \frac{n}{N}\frac{\Delta \tau}{T_\text{s}}\right)}.
\end{aligned}
\end{equation}
\end{subequations}

To compare Eqs.~\eqref{eq:ISI_fs_b} and~\eqref{eq:ISI_fn_b}, we note that the delay difference is determined by the path length difference, $\Delta L$, which typically ranges from a few meters to tens of meters in real-world scenarios, as $\Delta \tau = \Delta L / c$, where $c$ is the speed of light in free space. For a symbol duration of $T_\text{s} = 1~\mu\text{s}$, this implies that $\Delta \tau$ is typically smaller than $T_\text{s}$, and hence the arguments of the sine functions in both $\Delta r_\text{s}$ and $\Delta r_n$ lie within the monotonic interval $[0, \pi/2]$. It follows from $n/N < 1$ that $\Delta r_\text{s}>\Delta r_n$. This result confirms the earlier intuition that narrow-band subcarriers in OFDM exhibit reduced sensitivity to multipath delay, thereby offering improved ISI robustness compared to WSC modulation. This advantage is particularly valuable in wireless environments with significant delay spreads.

\section{Benefit of the Prefix in OFDM}
\label{sec:appendix_B_prefix}
\begin{figure*}[h!]
  \centering
  \begin{subfigure}{\linewidth}
    \centering
    \includegraphics[width=\linewidth]{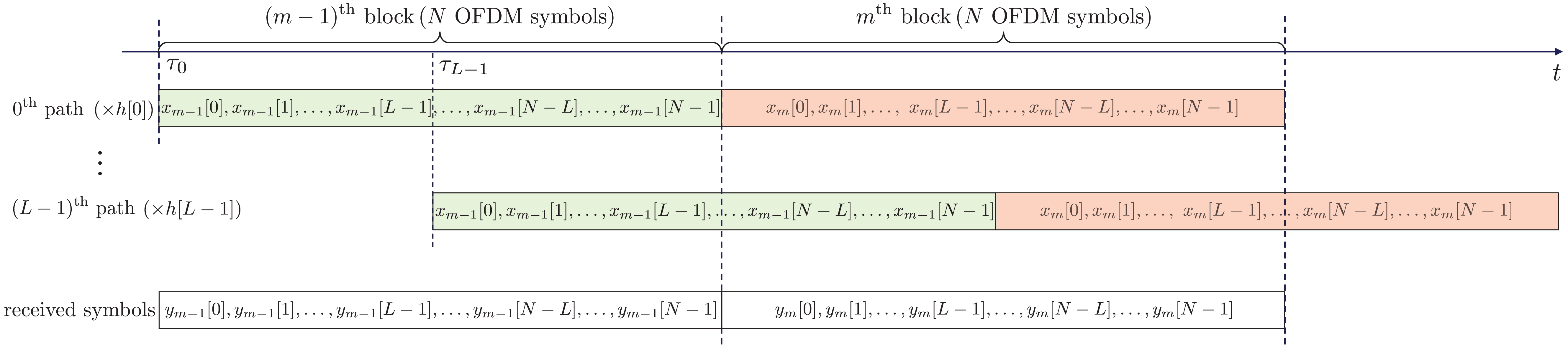}
    \captionsetup{skip=1pt}  
    \caption{}
    \label{fig:ISI_withoutCP}
  \end{subfigure}

  \begin{subfigure}{\linewidth}
    \centering
    \includegraphics[width=\linewidth]{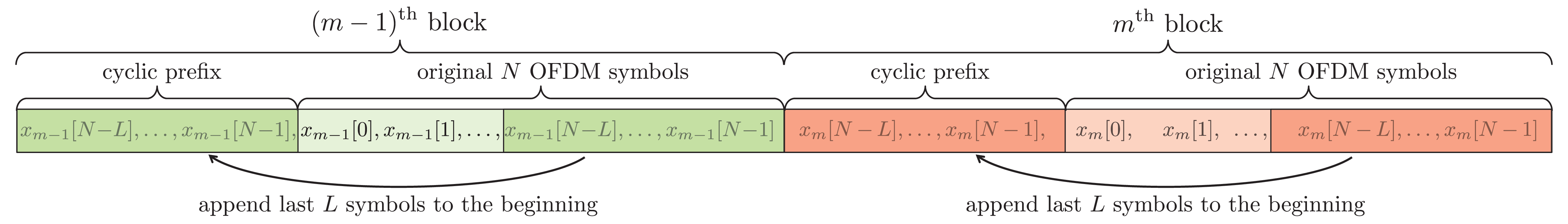}
    \captionsetup{skip=1pt}  
    \caption{}
    \label{fig:adding_CP}
  \end{subfigure}
  \vspace{0.1cm}
  
  \begin{subfigure}{\linewidth}
    \centering
    \includegraphics[width=\linewidth]{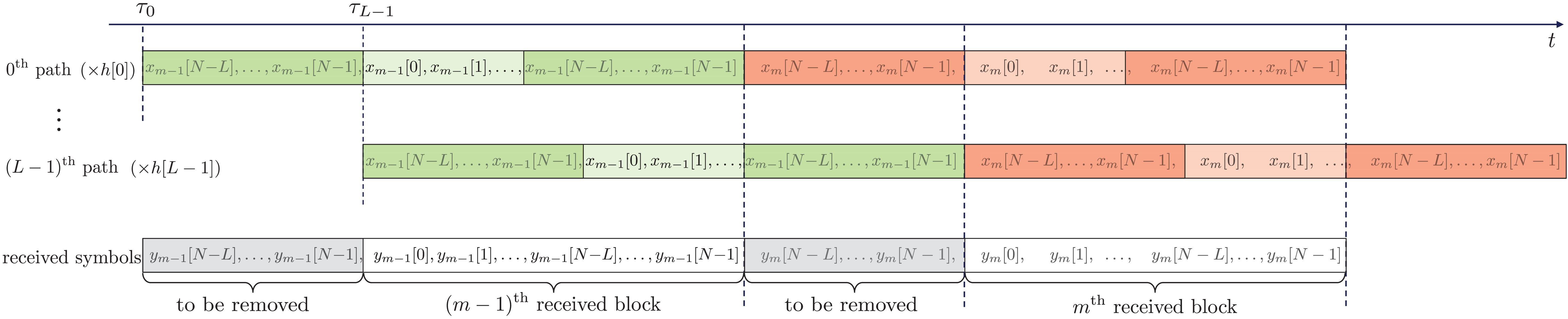}
    \captionsetup{skip=1pt}  
    \caption{}
    \label{fig:noISI_withCP}
  \end{subfigure}
  
  \caption{
Suppression of inter-symbol interference (ISI) due to multipath fading by the cyclic prefix (CP) technique in OFDM. 
(a)~ISI arising in the absence of cyclic prefix. The delay spread of the multipath channel causes symbols from the $(m{-}1)^\text{th}$ block to interfere with the $m^\text{th}$ block.
(b)~Construction of the CP sequence by copying the last $L$ symbols of each OFDM block and appending them to the beginning of the block before transmission.
(c)~ISI suppression with CP in place. The ISI caused by the block $m-1$ affects only the CP portion of the received signal, which is discarded at the receiver, ensuring that the remaining $N$ samples in the $m^\text{th}$ block are free from inter-block interference, to allow treating channel convolution as circular.
}
\label{fig:CP_illustration}
\end{figure*}

As discussed in Appendix~\ref{sec:appendix_A_narrowband}, the adoption of narrow-band subcarriers in OFDM reduces but does not fully suppress the ISI caused by multipath fading. To further mitigate ISI, a prefix\footnote{CP is widely adopted in practical systems due to its ability to directly converts linear convolution into circular convolution, enabling efficient one-tap DFT-based equalization. ZP retains linear convolution, but after discarding the first $L$ received samples, the remaining portion becomes numerically equivalent to a circular convolution with a zero-padded channel. Despite requiring more complex equalization, ZP is often preferred in energy-constrained systems.}—typically a CP, or alternatively a ZP—is inserted at the beginning of each OFDM block~\cite{goldsmith2005wireless}. This will be explained next with the help of Fig.~\ref{fig:CP_illustration}.

\pati{Adding Prefix for Mitigating the ISI}
Equation~\eqref{eq:y_m_RTFT_dis} implies that each received symbol $y_m[n]$ is affected by the preceding symbols, due to multipath delays, as illustrated in Fig.~\ref{fig:CP_illustration}(a). By partitioning the index $n$ into two distinct intervals, the received symbols in the $m^{\text{th}}$ OFDM block [Eq.~\eqref{eq:y_m_RTFT_dis}] can be reformulated as
\begin{subequations} \label{eq:y_m_ISI} 
\begin{align}
y_{m}[n] &= \sum_{l=0}^{n}h[l]\, x_{m}\left[n-l\right]
+ \sum_{l=n+1}^{L-1}h[l]\, x_{m-1}\left[N-l\right], \notag \\
&\quad\text{for } 0 \leq n < L - 1, 
\label{eq:y_m_ISI_a} \\[1ex]
y_{m}[n] &= \sum_{l=0}^{L-1}h[l]\, x_{m}\left[n-l\right], \notag \\
&\quad\text{for } L - 1 \leq n \leq N - 1. 
\label{eq:y_m_ISI_b}
\end{align}
\end{subequations}
In Eq.~\eqref{eq:y_m_ISI_a}, ISI in the $m^{\text{th}}$ block arises from both the current block and the preceding $(m-1)^{\text{th}}$ block, the latter contribution being referred to as inter-block interference (IBI), while in Eq.~\eqref{eq:y_m_ISI_b} ISI in a block stems solely from symbols within the same block.

To eliminate the ISI described in Eq.~\eqref{eq:y_m_ISI}, a CP\footnote{In this appendix, CP is used as an example to illustrate its role in eliminating ISI in conventional OFDM, and the equivalent effect of a ZP is explained Sec.\ref{sec:Demonstration_B}.} of length $L$ is appended to the beginning of each OFDM block. The CP of the $m^{\text{th}}$ OFDM block corresponds to the last $L$ symbols of the block. As illustrated in Fig.~\ref{fig:CP_illustration}(b), the operation of CP insertion, yields a new data sequence $\{x^{\text{cp}}_{m}[n]\}$ of length $N+L$, indexed by $-L \leq n \leq N-1$, and given by
\begin{equation} \label{eq:Tx_OFDM_CP}
\begin{aligned}
\{x^{\text{cp}}_{m}[-L], \dots,  x^{\text{cp}}_{m}[-1], x^{\text{cp}}_{m}[0],\dots, x^{\text{cp}}_{m}[N-1]\}=\\\{x_{m}[N-L], \dots,  x_{m}[N-1], x_{m}[0],\dots, x_{m}[N-1]\},
\end{aligned}
\end{equation}
which, using $[n]_N$ to denote $n$ mod $N$, can be compactly written as
\begin{subequations}\label{eq:Tx_OFDM_periodic}
\begin{equation} \label{eq:Tx_OFDM_periodic_a}
\begin{aligned}
x^{\text{cp}}_{m}[n]=x_{m}[n]_N,~~-L\leq n\leq N-1,
\end{aligned}
\end{equation}
implying that $x^{\text{cp}}_{m}[n]$ is a periodic extension of $x_m[n]$ with period $N$, and
\begin{equation} \label{eq:Tx_OFDM_periodic_b}
x^{\text{cp}}_{m}[n - l] = x_m[n - l]_N,\quad -L \leq n - l \leq N-1.
\end{equation}
\end{subequations}
According to Eq.~\eqref{eq:y_m_RTFT_dis}, the corresponding received symbols of $\{x^{\text{cp}}_{m}[n]\}$ are given by
\begin{equation} \label{eq:y_mp_dis}
\begin{aligned}
y^{\text{cp}}_{m}[n]=\sum_{l=0}^{L-1}h[l] x^{\text{cp}}_{m}\left[n-l\right],~~-L\leq n\leq N-1,
\end{aligned}
\end{equation}
which, similar to the reformation in Eq.~\eqref{eq:y_m_ISI}, Eq.~\eqref{eq:y_mp_dis}, can be reformulated as
\begin{subequations} \label{eq:y_mp_ISI} 
\begin{align}
y^{\text{cp}}_{m}[n]  &= \sum_{l=0}^{n}h[l]\, x^{\text{cp}}_{m}\left[n-l\right]+ \sum_{l=n+1}^{L-1}h[l]\, x^{\text{cp}}_{m-1}\left[N-l\right], \notag \\
&\quad\text{for } -L \leq n < 0, 
\label{eq:y_mp_ISI_a} \\[1ex]
y^{\text{cp}}_{m}[n]&=\sum_{l=0}^{L-1}h[l] x^{\text{cp}}_{m}\left[n-l\right], \notag \\
&\quad\text{for } 0\leq n\leq N-1. 
\label{eq:y_mp_ISI_b}
\end{align}
\end{subequations}
As observed in Eq.~\eqref{eq:y_mp_ISI_a}, the first $L$ samples of $\{y^{\text{cp}}_{m}[n]\}$ are corrupted by ISI originating from the last $L$ symbols of the previous block $\{x_{m-1}[n]\}$, as illustrated in Fig.~\ref{fig:CP_illustration}(c). They will therefore be discarded. The remaining $N$ symbols of $\{y^{\text{cp}}_{m}[n]\}$, given by Eq.~\eqref{eq:y_mp_ISI_b}, are only interfered with the symbols in the same block. Substituting Eq.~\eqref{eq:Tx_OFDM_periodic_b} transforms this relation into
\begin{equation} \label{eq:y_m_circular_conv_1}
\begin{aligned}
y^{\text{cp}}_{m}[n]=\sum_{l=0}^{L-1}h[l] x_{m}\left[n-l\right]_N,\quad 0 \leq n \leq N - 1,
\end{aligned}
\end{equation}
which implies that the last $N$ symbols of $\{y^{\text{cp}}_{m}[n]\}$ form a circular convolution\footnote{The circular convolution is defined as $x[n] \circledast h[n] = \sum_{l=0}^{N - 1} h[l]x[n - l]_N$. When $x[n]$ is of length $N$ and $h[n]$ is of length $L < N$, $h[n]$ is zero-padded to length $N$ before applying the circular convolution.} of the transmitted sequence $x_m[n]$ and the channel impulse response $h[l]$. Consequently, after discarding the first $L$ samples, the received signal in the $m^{\text{th}}$ OFDM block can be expressed as 

\begin{equation} \label{eq:y_m_circular_conv_2}
\begin{aligned}
y^\prime_{m}[n] = h[l] \circledast x_m[n],\quad 0 \leq n \leq N - 1.
\end{aligned}
\end{equation}
In discrete time, the circular convolution corresponds to multiplication in the frequency domain. Performing DFT to both side of Eq.~\eqref{eq:y_m_circular_conv_2} yields then
\begin{equation} \label{eq:Ym_DFT_ym}
\begin{aligned}
Y^\prime_m[i]&=\text{DFT}\{y^\prime_m[n]\}\\
&=\text{DFT}\{h[l] \circledast x_m[n]\}\\
&=H[i]X_m[i],\quad 0 \leq i \leq N - 1,
\end{aligned}
\end{equation}
which implies that the transmitted symbols $\{X_m[i]\}$ can be exactly recovered via the trivial operation $X_m[i]=Y_m[i]/H[i]$.

\begin{figure}[!h]
    \centering
    \begin{subfigure}[t]{0.48\linewidth}
        \includegraphics[width=\linewidth]{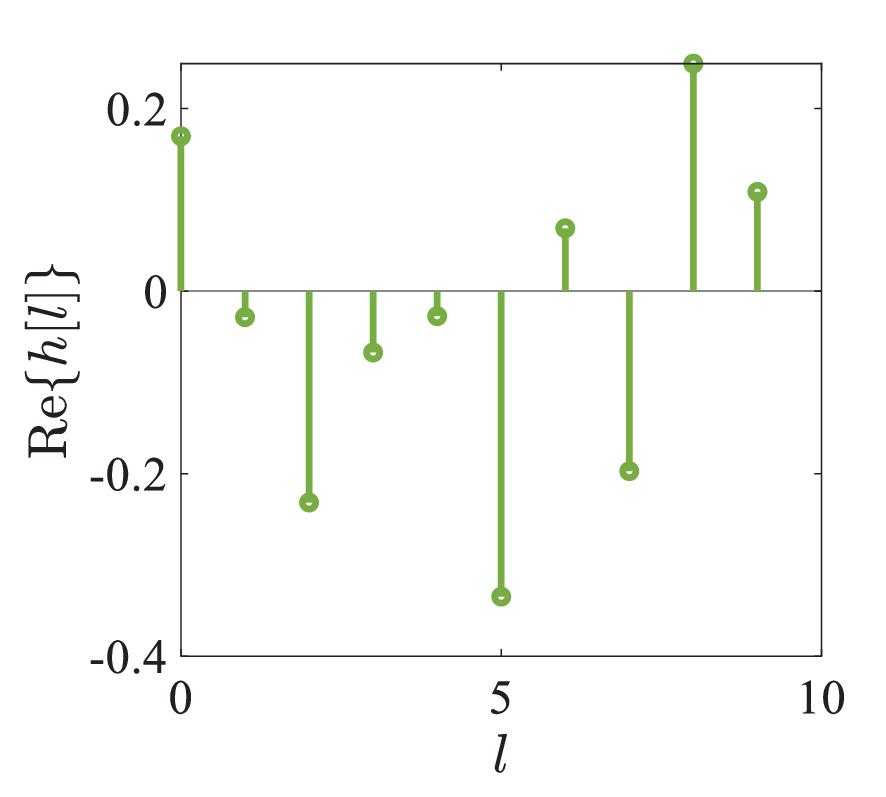}
        \captionsetup{skip=1pt}  
        \caption{}
        \label{fig:channel_response}
    \end{subfigure}
    \begin{subfigure}[t]{0.48\linewidth}
        \includegraphics[width=\linewidth]{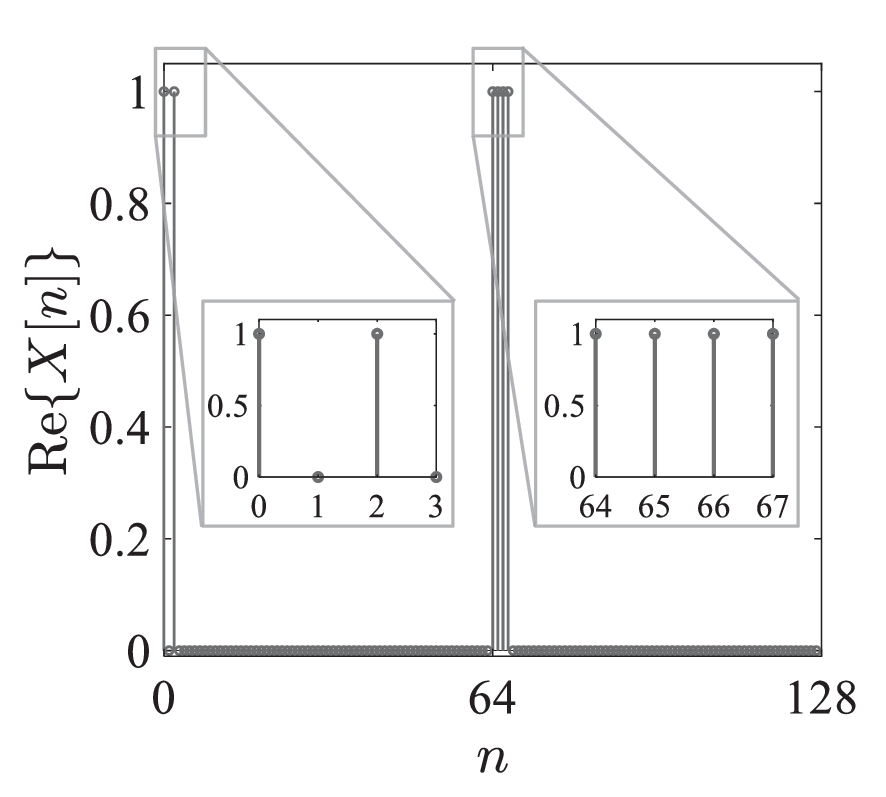}
        \captionsetup{skip=1pt}  
        \caption{}
        \label{fig:transmitted_X}
    \end{subfigure}
    
    \begin{subfigure}[t]{0.48\linewidth}
        \includegraphics[width=\linewidth]{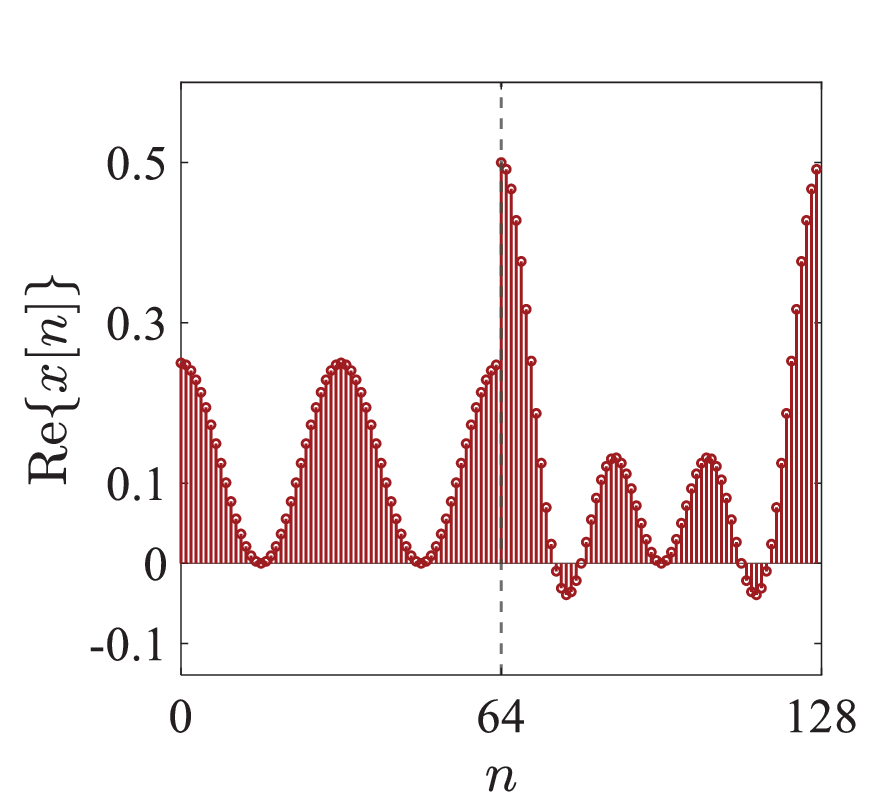}
        \captionsetup{skip=1pt}  
        \caption{}
        \label{fig:x_without_CP}
    \end{subfigure}
    \hspace{0.01cm}
    \begin{subfigure}[t]{0.48\linewidth}
        \includegraphics[width=\linewidth]{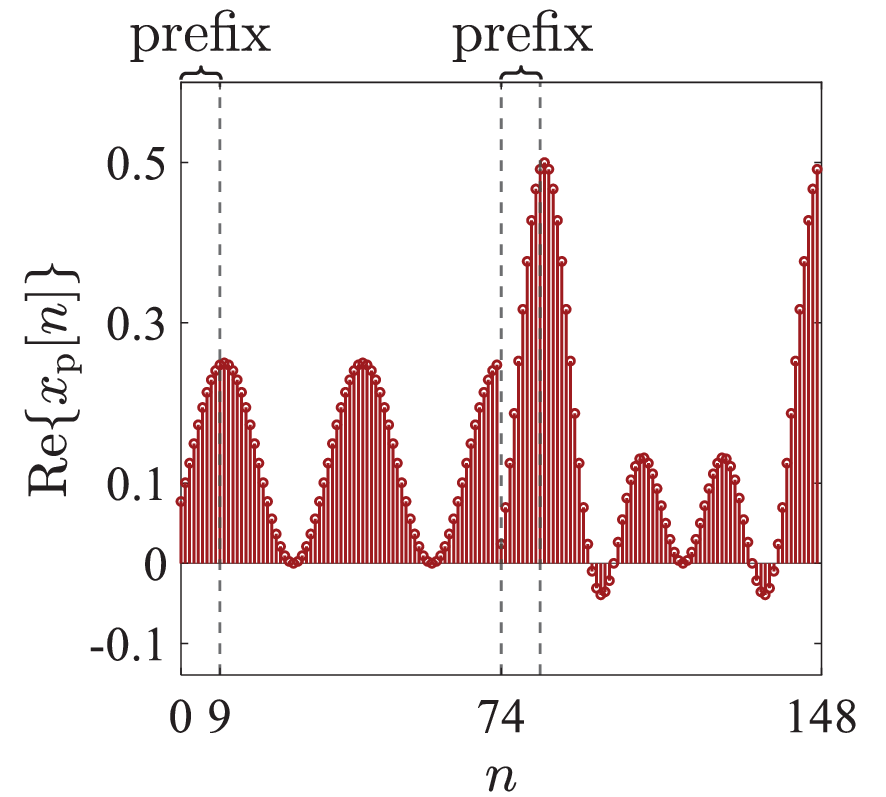}
        \captionsetup{skip=1pt}  
        \caption{}
        \label{fig:x_with_CP}
    \end{subfigure}
    \begin{subfigure}[t]{0.48\linewidth}
        \includegraphics[width=\linewidth]{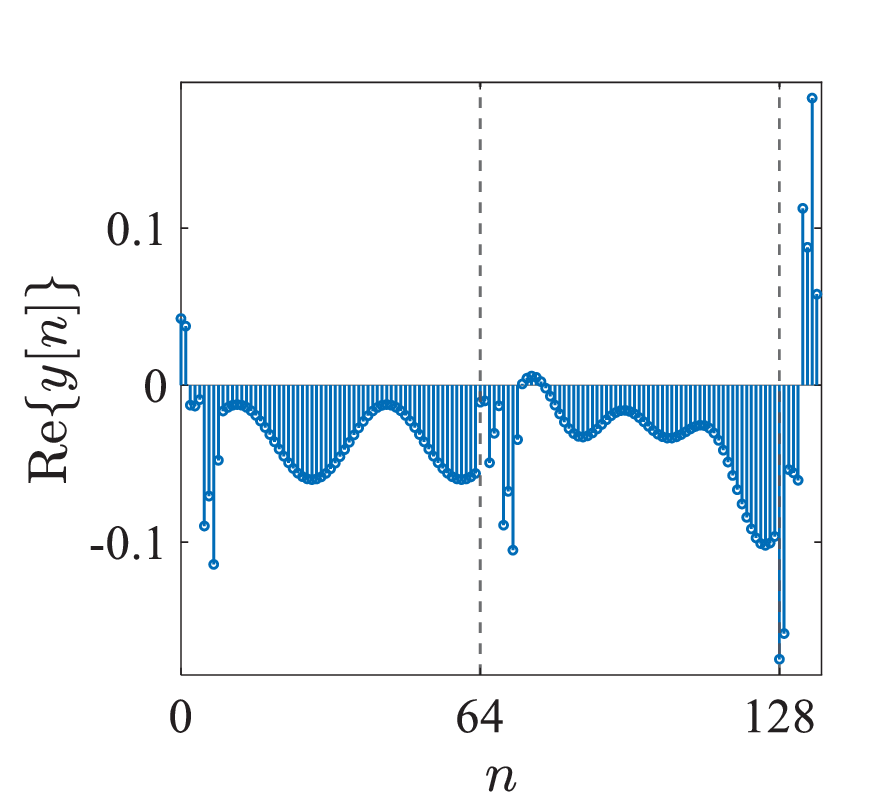}
        \captionsetup{skip=1pt}  
        \caption{}
        \label{fig:y_without_CP}
    \end{subfigure}
    \hspace{0.01cm}
    \begin{subfigure}[t]{0.48\linewidth}
        \includegraphics[width=\linewidth]{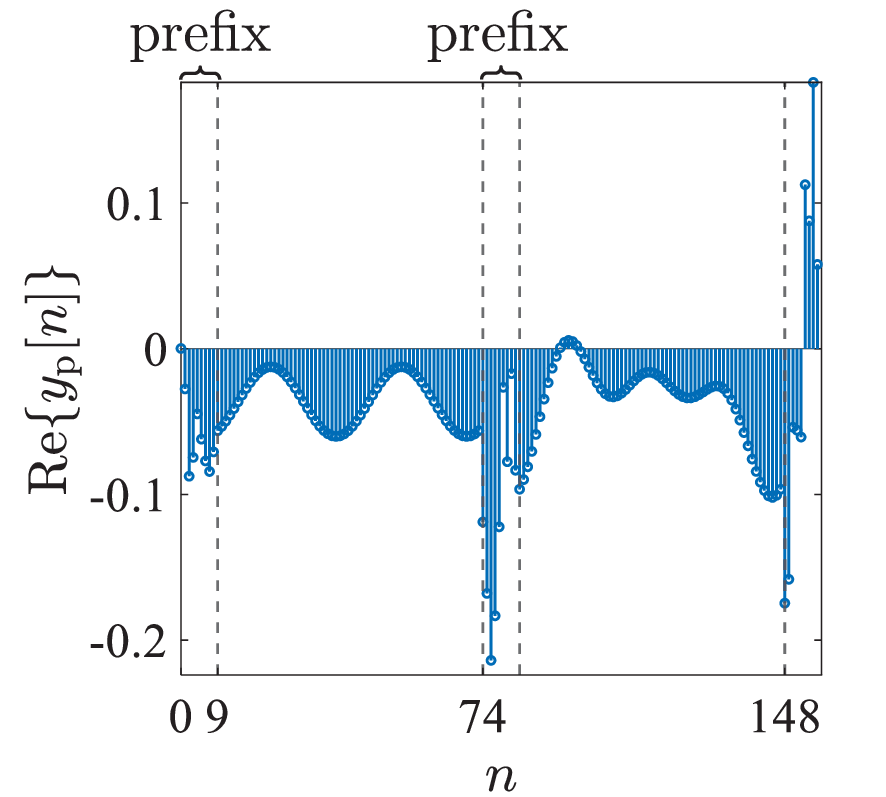}
        \captionsetup{skip=1pt}  
        \caption{}
        \label{fig:y_with_CP}
    \end{subfigure}
    \caption{Mitigation of ISI by the addition of a cyclic prefix (CP) in OFDM with $N=64$. 
    (a)~Complex channel impulse response (real part), $\gamma_l$, for each multi path component [Eq.~\eqref{eq:channel_response}]. 
    (b)~Transmitted symbols in two consecutive OFDM blocks ($\{X_{m-1}[n]\}=\{1,0,1,0,0,0,\dots,0\},~ \{X_{m}[n]\}=\{1,1,1,1,0,0,\dots,0\}$). 
    (c)~DFT (real part) of the data symbols in (b) [terms in Eq.~\eqref{eq:OFDM_disc}]. 
    (d)~OFDM symbols including a cyclic prefix [A portion (of length $L$) at the end of each block is copied and pasted to the front of this block]. 
    (e)~Received symbols corresponding to (c) through the channel characterized by (a). 
    (f)~Same as (e) but with correspondence to (d).}
    \label{fig:prefix_conventional_OFDM}
\end{figure}

Figure~\ref{fig:prefix_conventional_OFDM} illustrates the impact of adding a CP in mitigating ISI in an OFDM system with $N = 64$. The channel impulse response, shown in Fig.~\ref{fig:prefix_conventional_OFDM}(a), contains multiple delayed components with varying complex gains, representing a typical frequency-selective fading channel.  The input data symbols in two consecutive OFDM blocks, $\{X_{m-1}[n]\}$  and $ \{X_m[n]\}$, are depicted in Fig.~\ref{fig:prefix_conventional_OFDM}(b). Their IDFT, yielding the OFDM signals without a prefix, are shown in Fig.~\ref{fig:prefix_conventional_OFDM}(c). Figure~\ref{fig:prefix_conventional_OFDM}(d) presents the resulting signal after appending a CP to each block, where the last $L$ samples of each OFDM symbol are prepended to the beginning. The received signals after transmission through the channel are shown in Fig.~\ref{fig:prefix_conventional_OFDM}(e) and Fig.~\ref{fig:prefix_conventional_OFDM}(f), corresponding to the cases without and with a CP, respectively. It is evident from these figures that the CP effectively eliminates ISI, corresponding to the highly irregular part of the received symbols, across adjacent blocks by preserving circular convolution, thereby allowing straightforward frequency-domain equalization.

\setlength{\abovecaptionskip}{0pt}

\section{Microwave Implementation of RTFT}
\label{sec:appendix_C_RTFT}
\begin{figure}[h!]
\centering
\includegraphics[width=8cm]{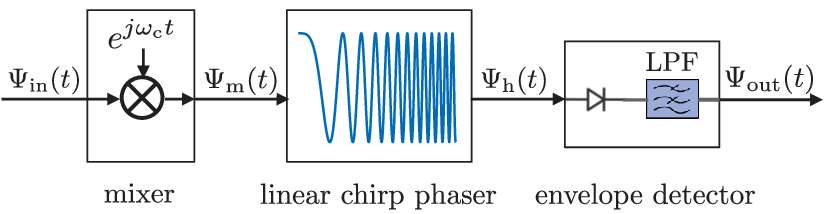}
\vspace{2mm}
\captionsetup{font=small}
\caption{Microwave implementation of an RT-FT processor, consisting of a mixer, a linear phaser (all-pass filter with linear group delay versus frequency) and an envelope detector~\cite{caloz2013analog}.}
\label{fig:Phaser_RTFT}
\end{figure}

\pati{RTFT Implementation using a Phaser}
Following the microwave analog signal processing framework proposed in~\cite{caloz2013analog}, Fig.~\ref{fig:Phaser_RTFT} presents the block diagram of a microwave implementation of RTFT, whose key component is a linear-chirp phaser.

A linear-chirp phaser is an all-pass filter with a group delay $\tau$ that varies linearly with frequency or, equivalently, with a quadratic phase response. Its transfer function can be expressed as
\begin{equation}\label{eq:TF_phaser}
H(\omega)=e^{j\phi(\omega)}=e^{j\phi_0}e^{j\phi_1(\omega-\omega_\text{c})}e^{j\frac{\phi_2}{2}(\omega-\omega_\text{c})^2},
\end{equation}
where $\omega_\text{c}$ is the center frequency, and $\phi_0$, $\phi_1$, and $\phi_2$ are the phase (rad), group delay (s) and group delay dispersion (s$^2$/rad), respectively. Taking the inverse Fourier transform of $H(\omega)$ yields the impulse response
\begin{subequations}
\begin{equation}\label{eq:IR_phaser}
\begin{split}
h(t)&=\int_{-\infty}^{+\infty}H(\omega)e^{j\omega t}d\omega\\
&=\int_{-\infty}^{+\infty}e^{j\phi_0}e^{j\phi_1\left(\omega-\omega_\text{c}\right)}e^{j\frac{\phi_2}{2}(\omega- \omega_\text{c})^2}e^{j\omega t}d\omega\\
&=\gamma e^{-j\frac{(\phi_1-\omega_\text{c}\phi_2)t}{\phi_2}}e^{-j\frac{t^2}{2\phi_2}},
\end{split}
\end{equation}
with
\begin{equation}\label{eq:phaser_gamma}
\gamma=\sqrt{\frac{2\pi}{\phi_2}}e^{j\left(\frac{\pi}{4}+\phi_0+\phi_1\omega_\text{c}-\frac{\phi_1^2}{\phi_2}+\frac{\phi_2\omega_\text{c}^2}{2}-\phi_2\omega_\text{c}^2\right)}.
\end{equation}
\end{subequations}

As illustrated in Fig.~\ref{fig:Phaser_RTFT}, the baseband input signal $\Psi_\text{in}(t)$ is first upconverted to the microwave frequency band, centered at $\omega_\text{c}$, resulting in the modulated signal $\Psi_\text{m}(t) = \Psi_\text{in}(t)e^{j\omega_\text{c}t}$. This signal is then passed through the linear-chirp phaser, producing an output $\Psi_\text{h}(t)$ that corresponds to the convolution between $\Psi_\text{m}(t)$ and the phaser impulse response $h(t)$, viz.,
\begin{equation}\label{eq:phaser_output}
\begin{split}
\Psi_\text{h}(t)&=\Psi_\text{m}(t)*h(t)\\
&=\int_{-\infty}^{+\infty}\Psi_\text{m}(\tau) h(t-\tau)d\tau\\
&=\int_{-\infty}^{+\infty}\left[\Psi_\text{in}(\tau) e^{j\omega_\text{c}\tau}\right]\left[\gamma e^{-j\frac{(\phi_1-\omega_\text{c}\phi_2)(t-\tau)}{\phi_2}}e^{-j\frac{(t-\tau)^2}{2\phi_2}}\right]d\tau\\
&=\gamma e^{-j\frac{(\phi_1-\omega_\text{c}\phi_2)t}{\phi_2}}e^{-j\frac{t^2}{2\phi_2}}\int_{-\infty}^{+\infty}\Psi_\text{in}(\tau) e^{j\frac{\phi_1+t}{\phi_2}\tau}e^{-j\frac{\tau^2}{2\phi_2}}d\tau.
\end{split}
\end{equation}
In the last equality, it is evident that the first exponential term in the integrand, $e^{j\frac{\phi_1 + t}{\phi_2} \tau}$, resembles the kernel of a Fourier transform, with a time-dependent frequency defined by $\omega(t) = -(\phi_1 + t)/\phi_2$. For the integral in Eq.~\eqref{eq:phaser_output} to precisely represent the Fourier transform of $\Psi_\text{in}$, the second exponential term, $e^{-j\frac{\tau^2}{2\phi_2}}$, must be unity over the duration of the input signal. This condition is approximately satisfied if the phase variation introduced by the quadratic term is negligible, i.e., $\tau_\text{max}^2/(2|\phi_2|)\ll \pi$, or equivalently,
\begin{equation}\label{eq:far_field}
    \frac{T_\text{in}^2}{2\pi |\phi_2|}\ll 1,
\end{equation}
where $T_\text{in}$ represents the duration of the input signal. Equation~\eqref{eq:far_field} is referred to as the  \textit{time-domain far-field condition}. It imposes a constraint on the allowable input signal duration and group delay dispersion parameter $\phi_2$, under which the system can accurately implement RTFT.

Unfortunately, the condition on the phaser parameter $\phi_2$ derived in Eq.~\eqref{eq:transmitter_phi2} for generating the OFDM signal is not sufficient to guarantee the condition in Eq.~\eqref{eq:far_field}. Since the input signal duration within each OFDM block is $T_\text{in}=T_0$, substituting this along with the value of $\phi_2$ in Eq.~\eqref{eq:transmitter_phi2} into the left side of Eq.~\eqref{eq:far_field} yields ${T_\text{in}^2}/({2\pi |\phi_2|})=N$, where $N$ is the number of subcarriers in the OFDM system. The resulting condition, $N\ll1$, is clearly violated, implying that the second exponential term in the integrand of Eq.~\eqref{eq:phaser_output} cannot be neglected. In fact, only a small portion of the input signal duration, specifically for $\tau\in(0, \sqrt{N}T_\text{s})$ (with $N\ll1$), satisfies the condition under which the second exponential term becomes negligible.

\begin{figure*}[h]
  \centering
    \includegraphics[width=0.8\linewidth]{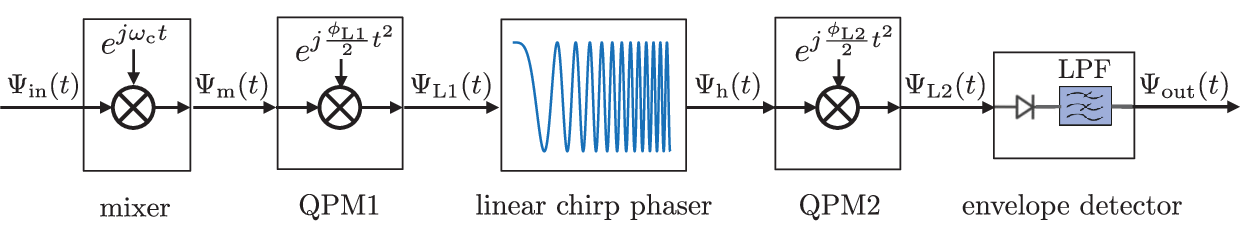}
    \captionsetup{skip=1pt}  
  \caption{Enhanced microwave implementation of an RT-FT, incorporating two quadratic phase modulators (QPM1 and QPM2)~\cite{kolner1989temporal} in addition to the standard mixer, linear chirp phaser and envelope detector. QPM1 and QPM2 eliminate the need for the far-field condition [Eq.~\eqref{eq:far_field}] and enable accurate frequency-to-time mapping over the full input duration.
}
\label{fig:Phaser_with_QPM}
\end{figure*}

To overcome the limitation in Eq.~\eqref{eq:far_field} for eliminating the unwanted quadratic phase term in Eq.~\eqref{eq:phaser_output}, a quadratic phase modulator (QPM), also referred to as a time lens~\cite{kolner1989temporal,salem2008optical,ojeda2009periodic,zhang2014ultrafast}, may be incorporated before and after the phaser, as shown in Fig.~\ref{fig:Phaser_with_QPM}. This QPM introduces a quadratic phase modulation of the form $e^{j\Phi_l t^2/2}$, where $\Phi_l = 1/\phi_2$ that precisely cancels the quadratic phase term in the integrand of Eq.~\eqref{eq:phaser_output}, thereby providing exact Fourier transformation regardless of the input duration. Fortunately, the additional QPM's are commercially available and do not consume significant power~\cite{ojeda2009periodic}.

In the following, we demonstrate how the use of quadratic phase modulation eliminates the far-field condition in Eq.~\eqref{eq:far_field} and ensures faithful frequency-to-time mapping over the entire OFDM block. As illustrated in Fig.~\ref{fig:Phaser_with_QPM}, the signal at the output of QPM1 output is
\begin{equation}\label{eq:output_QPM1}
\begin{aligned}
\Psi_{\text{L}1}(t)=\Psi_\text{in}(t) e^{j\omega_\text{c}t}e^{j\frac{\phi_{\text{L}1}}{2}t^2}.
\end{aligned}
\end{equation}
This signal passes then through the phaser, and emerges from it as the convolution between $\Psi_{\text{L}1}(t)$ and the phaser impulse response $h(t)$ defined in Eq.~\eqref{eq:IR_phaser}, viz.,
\begin{subequations}
\begin{equation}\label{eq:output_phaser_a}
\begin{aligned}
\Psi_\text{h}(t)&=\Psi_{\text{L}1}(t)*h(t)\\
&=\int_{-\infty}^{+\infty}\Psi_{\text{L}1}(\tau) h(t-\tau)d\tau\\
&=\int_{-\infty}^{+\infty}\left[\Psi_\text{in}(\tau) e^{j\omega_\text{c}\tau}e^{j\frac{\phi_{\text{L}1}}{2}\tau^2}\right]\left[\gamma e^{-j\frac{\left(\phi_1-\omega_\text{c}\phi_2\right)\left(t-\tau\right)}{\phi_2}}e^{-j\frac{\left(t-\tau\right)^2}{2\phi_2}}\right]d\tau\\
&=\gamma e^{-j\frac{\left(\phi_1-\omega_\text{c}\phi_2\right)t}{\phi_2}}e^{-j\frac{t^2}{2\phi_2}}\\&~~~\int_{-\infty}^{+\infty}\Psi_\text{in}(\tau) e^{j\frac{\phi_1+t}{\phi_2}\tau}e^{j\frac{1}{2}\left(\phi_{\text{L}1}-\frac{1}{\phi_2}\right)\tau^2}d\tau,
\end{aligned}
\end{equation}
which, with the substitution of $\phi_{\text{L}1}=1/\phi_2$, becomes
\begin{equation}\label{eq:output_phaser_b}
\begin{aligned}
\Psi_\text{h}(t)=\gamma e^{-j\frac{\left(\phi_1-\omega_\text{c}\phi_2\right)t}{\phi_2}}e^{-j\frac{t^2}{2\phi_2}}\int_{-\infty}^{+\infty}\Psi_\text{in}(\tau) e^{j\frac{\phi_1+t}{\phi_2}\tau}d\tau.
\end{aligned}
\end{equation}
\end{subequations}
We see in this expression that the spurious far-field term has been eliminated from the integral, which has then become a perfect RT-FT operation. The quadratic phase term outside the integral can then be eliminated by a second phase modulator, QPM2, as
\begin{subequations}
\begin{equation}\label{eq:output_QPM2_a}
\begin{aligned}
\Psi_{\text{L}2}(t)&=\Psi_\text{h}(t)e^{j\frac{\phi_{\text{L}2}}{2}t^2}\\
&=\gamma e^{-j\frac{\left(\phi_1-\omega_\text{c}\phi_2\right)t}{\phi_2}}e^{j\frac{1}{2}\left(\phi_{\text{L}2}-\frac{1}{\phi_2}\right)t^2}\int_{-\infty}^{+\infty}\Psi_\text{in}(\tau) e^{j\frac{\phi_1+t}{\phi_2}\tau}d\tau,
\end{aligned}
\end{equation}
which, upon setting $\phi_{\text{L}2} = 1/\phi_2$, becomes
\begin{equation}\label{eq:output_QPM2_b}
\begin{aligned}
\Psi_{\text{L}2}(t)=\gamma e^{-j\frac{\left(\phi_1-\omega_\text{c}\phi_2\right)t}{\phi_2}}\int_{-\infty}^{+\infty}\Psi_\text{in}(\tau) e^{j\frac{\phi_1+t}{\phi_2}\tau}d\tau,
\end{aligned}
\end{equation}
\end{subequations}
where the time-varying term outside the integral can be easily removed by envelope detection, leading to the baseband  signal
\begin{equation}\label{eq:RTFT_out}
\Psi_\text{out}\left[\omega(t)\right]=\sqrt{\frac{2\pi}{|\phi_2|}}\int_{-\infty}^{+\infty}\Psi_\text{in}(\tau)e^{j\omega(t)\tau}d\tau,
\end{equation}
where $\omega(t)$ is the instantaneous frequency given by Eq.~\eqref{eq:RTFT_omega_t}. This signal is the desired real-time Fourier transform in Eq.~\eqref{eq:RTFT_GVD}.

\section{Implementation Feasibility}
\label{sec:appendix_E_feasibility}
\begin{figure}[h!]
\centering
\includegraphics[width=6cm]{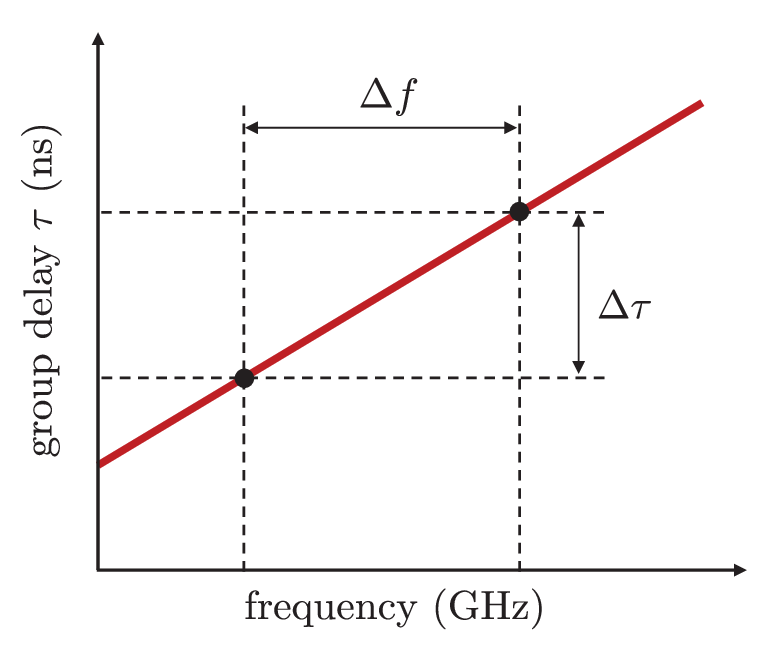}
\captionsetup{font=small}
\caption{Group delay response of a microwave linear-chirp phaser.}
\label{fig:group_delay}
\end{figure}

\pati {Feasibility Analysis of the Required Phaser Parameters}
Based on the phaser transfer function provided in Eq.~\eqref{eq:TF_phaser}, the group delay of the phaser~\cite{caloz2013analog} is
\begin{equation}\label{eq:group_delay}
\begin{aligned}
\tau(\omega)&=-\frac{\partial \phi(\omega)}{\partial \omega}\\
&=-\phi_1-\phi_2\left(\omega-\omega_\text{c}\right),
\end{aligned}
\end{equation}
This expression describes a group delay that varies linearly with frequency, as illustrated in Fig.~\ref{fig:group_delay} and corresponding to a microwave-band phaser of the type demonstrated in previous works~\cite{gupta2010group,zhang2012synthesis}.

From Eq.~\eqref{eq:group_delay}, it is evident that $-\phi_1$ represents the group delay at the center frequency $\omega = \omega_\text{c}$. For this value to be physically realizable, $\phi_1$ must be on the order of nanoseconds. Moreover, according to Eq.~\eqref{eq:transmitter_phi1}, on must have $\phi_1 = -T_0 = -N T_\text{s}$, where $N$ is the number of subcarriers in a single OFDM block. To ensure this condition on $\phi_1$ within a practical range, the symbol duration $T_\text{s}$ must again be on the order of $1~\text{ns}$.

Let us find now the condition on the parameter $\phi_2$. From either Eq.~\eqref{eq:group_delay} or from the slope of the linear group delay curve in Fig.~\ref{fig:group_delay}, the relationship between the group delay swing $\Delta \tau$ and $\phi_2$ is given by
\begin{equation}\label{eq:phi2_tau}
\begin{aligned}
-\phi_2=\frac{\Delta \tau}{\Delta \omega}=\frac{\Delta \tau}{2\pi\Delta f},
\end{aligned}
\end{equation}
indicating that practical values of $\phi_2$ are typically on the order of $1~\text{ns}/(\text{GHz}\cdot\text{radian})$ or, equivalently, $1~\text{ns}^2/\text{radian}$. Furthermore, the required value of $\phi_2$ at the transmitter and receiver, denoted by $\phi_{\text{tr},2}$ and $\phi_{\text{re},2}$ in Eqs.~\eqref{eq:transmitter_phi2} and \eqref{eq:receiver_phi2}, must satisfy 
\begin{equation}\label{eq:phi2_required}
\begin{aligned}
|\phi_2|=\frac{T_0 T_\text{s}}{2\pi}=\frac{NT_\text{s}^2}{2\pi},
\end{aligned}
\end{equation}
which again implies that $T_\text{s}$ must be on the order of $1~\text{ns}$ (or less), to keep $\phi_2$ within a practical range.

The above requirements on the symbol duration are not only theoretically reasonable, but correspond to commercial OFDM communication standards, such 5G systems operating around the microwave frequency of 28~GHz~\cite{IEEE80216} ($T_\textrm{s}<1$~ns) or 6G systems operating around the millimeter-wave frequency of 60~GHz~\cite{IEEE802153c,IEEE80211ad,IEEE80211ay} ($T_\textrm{s}<0.5$~ns) or THz frequencies in the range of 252~GHz to 325~GHz~\cite{IEEE802153d} ($T_\textrm{s}<0.02$~ns).

An inherent limitation of the proposed analog OFDM scheme its restricted realizability of larger group delay dispersion~\cite{gupta2010group} for operation at frequencies below the aforementioned systems, e.g., near 5~GHz. Indeed, at such frequencies, increasing $\phi_2$ often leads to trade-offs, such as higher insertion loss, increased physical footprint and reduced fabrication tolerance. According to Eq.~\eqref{eq:phi2_required}, the required group delay dispersion for realizing analog OFDM is directly proportional to both the number of subcarriers $N$ and the symbol duration $T_\text{s}$. Therefore, in order to support a larger number of subcarriers $N$, a proportionally larger $\phi_2$ is needed--something that is difficult to achieve with current technology.

Several strategies have been proposed to  overcome this limitation and enhance the effective $\phi_2$. A promising approach involves using a feedback-loop architecture, where the signal is reinjected multiple times into the same phaser, with amplification to compensate for loss~\cite{nikfal2011increased}. This technique achieves resolution enhancement proportional to the number of loops. Alternatively, larger group delay dispersion can be achieved using systems with greater delay swings over comparably narrower bandwidths~\cite{gupta2012crlh,zhang2012synthesis}. Fortunately, given the current trend for higher-frequency systems in next-generation communication system, such developments might become unnecessary, the currently available phasers providing practically suitable responses in the related frequency ranges.

\end{appendices}




%
\bibliographystyle{IEEEtran}
\bibliography{2025_03_GVD_OFDM}

\end{document}